\documentclass[a4paper,11pt]{article}
\pdfoutput=1 % if your are submitting a pdflatex (i.e. if you have
\usepackage{jcappub} % for details on the use of the package, please
\usepackage[T1]{fontenc} % if needed
\usepackage{grffile}
\usepackage[nameinlink,capitalize]{cleveref}
\newcommand{\be}{\begin{equation}}
\newcommand{\ee}{\end{equation}}
\newcommand{\bea}{\begin{eqnarray}}
\newcommand{\eea}{\end{eqnarray}}
\def\dep{\delta p}
\def\cs2{c_s^2}

\title{\boldmath Relativistic effects in the large-scale structure with effective dark energy fluids}

\author[a,b]{Cristian Barrera-Hinojosa} %\note{Corresponding author.}
\author[a]{and Domenico Sapone}

\affiliation[a]{Departamento de F\'isica, FCFM, Universidad de Chile,\\ Blanco Encalada 2008, Santiago, Chile}
\affiliation[b]{Institute for Computational Cosmology, Department of Physics, Durham University,\\ Durham DH1 3LE, U.K.}

\emailAdd{cristian.g.barrera@durham.ac.uk}
\emailAdd{domenico.sapone@uchile.cl}

\abstract{We study the imprints of an effective dark energy fluid in the large scale structure of the universe through the observed angular power spectrum of galaxies in the relativistic regime. We adopt the phenomenological approach that introduces two parameters $\{Q,\eta\}$ at the level of linear perturbations and allow to take into account the modified clustering (or effective gravitational constant) and anisotropic stress appearing in models beyond $\Lambda$CDM. We characterize the effective dark energy fluid by an equation of state parameter $w=-0.95$ and various sound speed cases in the range $10^{-6}\leq \cs2\leq 1$, thus covering K-essence and quintessence cosmologies. We calculate the angular power spectra of standard and relativistic effects for these scenarios under the $\{Q,\eta\}$ parametrization, and we compare these relative to a fiducial $\Lambda$CDM cosmology. We find that, overall, deviations relative to $\Lambda$CDM are stronger at low redshift since the behavior of the dark energy fluid can mimic the cosmological constant during matter domination era but departs during dark energy domination. In particular, at $z=0.1$ the matter density fluctuations are suppressed by up to $\sim3\%$ for the quintessence-like case, while redshift-space distortions and Doppler effect can be enhanced by $\sim15\%$ at large scales for the lowest sound speed scenario. On the other hand, at $z=2$ we find deviations of up to $\sim5\%$ in gravitational lensing, whereas the Integrated Sachs-Wolfe effect can deviate up to $\sim17\%$. Furthermore, when considering an imperfect dark energy fluid scenario, we find that all effects are insensitive to the presence of anisotropic stress at low redshift, and only the Integrated Sachs-Wolfe effect can detect this feature at $z=2$ and very large scales.}

\begin{document}
\maketitle
\flushbottom

\section{Introduction} \label{sec:intro}

During the last decades, large galaxy surveys have revealed that matter is coherently distributed in the universe up to very large scales in the form of a complex network known as {\it the cosmic web}~\cite{Tegmark:2003}. This large scale structure consist mainly of dark matter halos, where galaxy clusters reside, long filaments connecting and feeding these regions, sheets (planar regions) and galactic voids. At the cosmological level, the structure formation is driven by the dynamical instability of self-gravitating dark matter, a process seeded by tiny initial primordial fluctuations. Then, the large scale structure of the universe encodes a wealth of information about its initial conditions, the fundamental properties of gravity and the matter-energy content. Several surveys have been planning to measure the distribution of galaxies (Euclid~\cite{Euclid}, LSST~\cite{LSST}, DESI~\cite{Aghamousa:2016zmz}) at high redshifts with unprecedented precision, and then a careful theoretical modelling is required to properly interpret the upcoming data.  

A key realization for understanding the data collected in galaxy surveys is that for measuring the distribution of galaxies in the universe we do not observe their real positions, but we rather infer them from the incoming photons that have been redshifted through the line of sight. The general procedure is to pixelize the observed distribution of galaxies in bins of redshift and solid angle and then to count the fluctuations in the number of galaxies across the sky. There are two main effects that influence the counting in numbers: 1) the difference between the volume elements constructed using the observed redshift and observed angles with respect to the real physical volume that the galaxies actually occupy; 2) the observed flux and redshift of these galaxies could differ from their intrinsic properties because photons are perturbed when traveling from source to observer. Then, it is clear that the observed galaxy number density contains additional contributions arising from the distortion in the observable quantities, in contrast to the standard description where galaxies simply trace the underlying matter distribution (up to a bias factor), and this can be naturally resolved if we construct theoretical predictions in terms of observable quantities.

Historically, the fact that observations are not directly made in position space but rather in redshift space was first realized by Kaiser~\cite{Kaiser} who introduced the concept of redshift-space distortion (RSD), with which the symmetry between the clustering of matter in real and redshift spaces is broken due to the peculiar motions of the sources. Then, the observed galaxy distribution is not only sensitive to the underlying dark matter but also to their peculiar velocities, which also introduces a Doppler effect component in the observed clustering of galaxies~\cite{Hamilton}. Later it was also shown that gravitational lensing also affects the observed matter distribution via the magnification bias~\cite{Broadhurst,Moessner} since in practice galaxy surveys are limited in magnitude. 

The previous observables are usually referred to as {\it standard effects} because they appear intrinsically in the observed clustering of matter. However, during the last decade various works have developed a general description of the large scale structure of the universe \cite{Bonvin-WL,Yoo,Yoo2,Bonvin-Durrer} which leads to naturally include a set of {\it relativistic effects} that account for the impact of General Relativity (GR) onto the photon propagation. These effects comprise gravitational redshift, the Sachs-Wolfe effect, Shapiro time-delay and the Integrated Sachs-Wolfe (ISW) effect, and share the common feature of being suppressed at sub-horizon scales with respect to the standard contributions. Nevertheless they allow to consistently interpret different observables such as the power spectra of galaxies, the Cosmic Microwave Background and gravitational waves signals~\cite{BonvinGW}. This can become particularly relevant at large scales since galaxy clustering in GR can become substantially different from its Newtonian approximation \cite{Bonvin,Bonvin-Hui-Gaztanaga}. On the other hand, since in alternative cosmological models the metric potentials and peculiar velocities respond different to the same underlying matter distribution compared to $\Lambda$CDM, the relativistic effects provide an interesting tool to study modified gravity and dark energy (DE) models.

In this work we investigate the imprints of effective DE fluids in the large scale structure through the observed angular power spectrum of matter including standard and relativistic effects. In particular, we consider DE and modified gravity models which can be described by a convenient phenomenological parametrization \cite{Amendola-Sapone, Sapone-Kunz, Sapone}. Adopting such framework we calculate the observed matter angular power spectrum for a wide class of models, which are effectively described by two new parameters which enter directly at the level of the Einstein Equations. As particular realizations, in this paper we consider K-essence and quintessence fluids~\cite{kessence,ArmendarizPicon:2000ah,Chiba:1999ka}. The former is not completely homogeneous in space but may have non-vanishing DE perturbations, namely it behaves as a fluid that might cluster at scales above a certain sound horizon which is set by a sound speed parameter. The fluctuations in the density field for a K-essence model may manifest superluminal behavior as shown in Ref.~\cite{Bonvin:2006vc,Babichev:2007dw}. However, in this paper we consider only constant sound speed over the entire expansion history of the Universe, restricting our analysis to cases with subluminal behaviors.

The rest of the paper is organized as follows: in \cref{sec:basic_equations} we fix our notation while introducing the main equations for the background cosmological model and the perturbation equations for general fluids. In \cref{sec:Q-eta-parametrization} we introduce the $\{Q,\eta\}$ parametrization, and the particular DE fluid model investigated in this work is presented in \cref{sec:DE-fluid-model}. In \cref{sec:matterdensity} we discuss the observed matter density contrast identifying standard and relativistic effects and we calculate its angular power spectrum when DE perturbations are taken into account, whose expressions are included in \cref{sec:appendix} and are valid for any model belonging to the $\{Q,\eta\}$ parametrization. We discuss the results for the K-essence and quintessence-like fluids in \cref{sec:results} modeled as perfect effective fluids, while the imprints of viscosity components in standard and relativistic effects are discussed in \cref{sec:viscosity}. 

\section{Background and perturbation equations}\label{sec:basic_equations}
We consider a spatially flat Friedmann-Lema\^itre-Robertson-Walker (FRLW) background universe with the presence of matter and a DE component. Then, the first Friedmann equation is
\begin{equation}\label{eq:Hubble}
H^2=H^2_0\left[\Omega_{m}a^{-3}+(1-\Omega_{m})a^{-3(1+\hat{w}(a))}\right]\,,
\end{equation}
where $a$ is the scale factor, $H\equiv a^{-1}{da}/{dt}$ is the Hubble parameter ($t$ represents cosmic time) and $\Omega_{m}\equiv8\pi G_N\rho_{m,0}/3H^2_0$ is the matter density fraction at the present time. In \cref{eq:Hubble}, $\hat{w}(a)$ is defined as 
\begin{equation}
\hat{w}(a) = \frac{1}{\ln a}\int_{1}^{a}\frac{w(a')}{a'}{\rm d}a' \nonumber
\end{equation}
where $w\equiv p/\rho$ corresponds to the equation of state (EoS) parameter for the DE component and controls the expansion history of the universe. From \cref{eq:Hubble} the comoving distance to an object at redshift $z$ is calculated as
\begin{equation}\label{eq:comoving-distance}
r(z)=\int_{0}^{z}\frac{dz'}{H(z')}\,.
\end{equation}
We also consider spatially flat cosmologies at the perturbed level. In the conformal Newtonian (or longitudinal) gauge the metric is \cite{Ma:1995ey}
\begin{equation}
ds^2=a^2(\tau)[-(1+2\Psi)d{\tau}^2+(1-2\Phi)\delta_{ij}dx^idx^j]\,,
\end{equation}
where $\Psi$ and $\Phi$ are the gauge-invariant scalar perturbations~\cite{Bardeen} (Bardeen potentials), $x^i$ are comoving coordinates and $\tau$ is the conformal time (and we will use overdots to denote derivative with respect to it). The equations for the matter sector follow from the conservation of the energy-momentum tensor $\nabla_\mu T^{\mu\nu}=0$, which at linear order result into the continuity and Euler equations~\cite{Ma:1995ey}
\begin{align}
    \dot\delta &= -(1+w) (\theta-\dot\Phi) - 3 \frac{\dot{a}}{a}\left(\frac{\dep}{\delta\rho}-w \right)\delta\,, \label{eq:delta} \\
\dot{\theta} &= -\frac{\dot{a}}{a}(1-3w) \theta-\frac{\dot{w}}{1+w}\theta+ \frac{\dep/\delta\rho}{1+w}k^2\delta+k^2(\Psi-\sigma)\,, \label{eq:v}
\end{align}
where $\delta=\delta\rho/\rho$ is the density contrast, $\theta\equiv ik^jV_j$ the velocity divergence, $\dep$ the pressure perturbation and $\sigma$ the anisotropic stress. Note that \cref{eq:delta} and \cref{eq:v} are valid for any non-interacting fluid components and need to be modified if these are coupled other than gravitationally. For instance, in interacting DE~\cite{Duniya:2015nva} and decaying dark matter~\cite{Audren:2014bca} models, \cref{eq:delta} and \cref{eq:v} can be violated due to the nongravitational DE--matter and matter--mater couplings. In this work we assume only two different species: cold dark matter, which is characterised by $w=\dep = \sigma=0$, and a non-interacting DE fluid which is only coupled to the former through gravity. For the latter these quantities are unknowns and they consist in three internal degrees of freedom. For the particular model discussed in this work (introduced later in~\cref{sec:DE-fluid-model}), our choice follows ref.~\cite{Kunz:2006wc}, in which we assume a constant in time EoS parameter and a pressure perturbation of the form
\be
\dep = \cs2\delta\rho + 3aH\left(\cs2-c_a^2\right)\frac{(1+w)\theta}{k^2}\,,
\ee
where $\cs2$ is the rest-frame sound speed at which fluctuations in the density field propagate and it controls the growth of perturbations in the DE fluid and $c_a^2\equiv\dot{p}/\dot{\rho}$ the adiabatic sound speed. This parametrization allows to cover a wide range of models such as quintessence ($\cs2=1$) and K-essence ($\cs2\neq1$) in an effective way~\cite{Sapone-Kunz}.

Let us briefly motivate the introduction of the effective $\{Q,\eta\}$ parameterization for DE models while we introduce the perturbation equations for the gravitational sector. In standard $\Lambda$CDM we assume that gravity is described by GR and the matter sector correspond to a perfect cosmic fluid (i.e. $\sigma=0$), so that, at the linear level, the $00$ and $ij$ components of Einstein equations in Fourier space are, respectively,
\begin{align}
-k^2\Phi&=4\pi G_Na^2\sum_i\rho_iD_i\,,\label{eq:Poisson}\\
\Psi&=\Phi\,,
\end{align}
where
\begin{equation}\label{comoving-gauge-density}
\rho_i D_i\equiv\delta\rho_i+3\mathcal{H}(\bar{\rho}_i+\bar{p}_i)\frac{V_i}{k}\,
\end{equation}
is the comoving density contrast of the $i$-th fluid component (often denoted by $\Delta$, but we reserve this for the observed matter overdensity variable introduced in \cref{sec:matterdensity}) and $\mathcal{H}=Ha$ is the conformal Hubble parameter. Naturally, in $\Lambda$CDM, the sum appearing in \cref{eq:Poisson} reduces to $\sum_i\rho_iD_i=\rho_m D_m$ as the cosmological constant is by definition completely homogeneous in space, and \cref{eq:Poisson} reduces to the standard Poisson equation sourced by matter overdensity.

\subsection{Effective parametrization of Dark Energy models}
\label{sec:Q-eta-parametrization}
In order to bring in modified gravity and DE models into our discussion we can introduce two functions $\{Q,\eta\}$ appearing directly at the level of the perturbation equations. Then, the 00 and $ij$ components of the Einstein equations can be phenomenologically recast as~\cite{Amendola-Sapone, Sapone-Kunz, Sapone}
\begin{align}
-k^2\Phi&=4\pi G_Na^2Q(a,k){\rho}_m D_m\,,\label{eq:00-Q}\\
\Psi&=[1+\eta(a,k)]\Phi\,,\label{eq:ij-eta}
\end{align}
where $D_m=\delta_m+3\mathcal{H}V/k$ is the comoving density contrast of matter. In this way, the role of $Q$ is to capture any extra contribution sourcing the gravitational potential $\Phi$ appearing beyond the standard $\Lambda$CDM cosmology. For instance, $Q\neq1$ can represent a possible clustering in the DE component (with $Q=1$ matching $\Lambda$), but in the context of modified gravity models this can also parametrize contributions from a fifth-force due to modifications of GR, which at the linear level effectively modifies Newton's constant as $G_{\text{eff}}(a,k)\equiv Q(a,k)G_N$~\cite{Koyama:2015vza}. The second model parameter $\eta$ allows to take into account a possible anisotropic stress that could arise due to differences in the scalar potentials $\Psi$ and $\Phi$ at some scales and/or redshifts. Again, this behavior can be due to viscosity components in the cosmic fluid but can also appear in alternatives to GR such as the Dvali-Gabadadze-Porrati (DGP) model~\cite{DGP,Koyama:2005kd} and $f(R)$ gravity~\cite{Hu-Sawicki,Saltas:2010tt}.

Since we are ultimately interested in calculating statistical quantities such as the power spectra of the different linear perturbations, it is convenient to introduce a set of transfer functions $\{T_D,T_V,T_\Psi,T_\Phi\}$ that relate their values at a given time to a single primordial metric perturbation $\Psi_{\text{in}}$ via
\begin{align}
D({\tau},\textbf{k})&=T_D({\tau},k)\Psi_{\text{in}}(\textbf{k})\,,\label{Td}\\
V({\tau},\textbf{k})&=T_V({\tau},k)\Psi_{\text{in}}(\textbf{k})\,,\label{Tv}\\
\Psi({\tau},\textbf{k})&=T_\Psi({\tau},k)\Psi_{\text{in}}(\textbf{k})\,,\label{Tpsi}\\
\Phi({\tau},\textbf{k})&=T_\Phi({\tau},k)\Psi_{\text{in}}(\textbf{k})\,.\label{Tphi}
\end{align}
Assuming the standard single-field inflationary scenario~\cite{Linde:1981}, the power spectrum of the primordial field $\Psi_{\text{in}}$ is characterized in terms of a spectral index $n_s$ and an amplitude $A_{s}$ as
\begin{equation}\label{eq:primordialPS}
k^3\langle\Psi_{\text{in}}(\textbf{k})\Psi^*_{\text{in}}(\textbf{k}')\rangle=(2\pi)^3A_{s}(k{\tau}_o)^{n_s-1}\delta(\textbf{k}-\textbf{k}')\,.
\end{equation}
Here, we have explicitly included the constant ${\tau}^{n_s-1}_o$ (the current comoving size of the horizon) in order to keep $A_{s}$ dimensionless for any value of $n_s$. Then, using \cref{Td}-\cref{Tphi}, the field equations \cref{eq:00-Q} and \cref{eq:ij-eta} can be written as
\begin{align}
T_D(z,k)&=-\frac{2a}{3\Omega_m}\left(\frac{k}{{H}_0}\right)^2\frac{T_\Psi}{Q(1+\eta)}\,,\label{TD-eq}\\
T_V(z,k)&=\frac{2a}{3\Omega_m}\frac{k\mathcal{H}}{{H}^2_0}\left[\left(1-\frac{a}{(1+\eta)^2}\frac{\partial\eta}{\partial a}\right)T_\Psi+\frac{a}{1+\eta}\frac{\partial{T}_\Psi}{\partial a}\right] \label{TV-eq}\,,\\
T_\Phi(z,k)&=\frac{T_\Psi}{(1+\eta)}\label{Tphi-eq}\,,
\end{align}
where we have explicitly included the $0i$ component in \cref{TV-eq} which relates the velocity and potentials. Then, for a given model (i.e. $\{Q,\eta\}$) this set of equations allows to calculate $T_D$, $T_V$ and $T_\Phi$ as a function of $T_\Psi$ only. Furthermore, within the linear theory we can decompose $T_\Psi$ in terms of a growth factor $G$ and a transfer function $T(k)$ at redshift zero as
\begin{equation}
T_\Psi(z,k)=G(a,k)T(k)\,.
\end{equation}
Here, $G\equiv a^{-1}{D_1(a)}/{D_1(a_\text{0})}$ describes the growth of structures in the universe relative to matter domination era, since in GR the linear growing mode evolves as $D_1(a)\sim a$ during that regime. Naturally, $G$ is sensitive to the presence of a DE component in the energy content of the universe, and we can conveniently encompass a wide class of models adopting the growth index formalism, in which we parametrize $G$ in terms of the growth index $\gamma$ as \cite{Linder}
\begin{equation}\label{definition-G}
G=G({a_\text{0}})\exp\left(\int_{a_\text{0}}^{a}da'\frac{\Omega_m(a')^\gamma-1}{a'}\right)\,,
\end{equation}
%
%\begin{equation}\label{definition-G}
%G(a,k)\equiv\frac{D_1}{a}=\exp\left(\int_{1}^{a}da'\frac{\Omega_m(a')^\gamma-1}{a'}\right)\,,
%\end{equation}
%
where
\begin{align}
\Omega_m(a)&=\frac{{\Omega_{m,0}a^{-3}}}{{(H/H_0)^2}}\,.\label{eq:Omega_a}
\end{align}
In $\Lambda$CDM the growth index takes the constant value $\gamma=6/11\sim0.545$, and then $G$ depends only on the scale factor, i.e. $G=G(a)$. In alternative models, however, $\gamma$ may be both scale and time dependent, and then it allows to parametrize possible deviations from the standard evolution of the perturbations in $\Lambda$CDM. The growth factor can either be modified at the background level by considering a different background cosmology, which leads to a different expansion history, or at the perturbative level, where the gravitational potentials might be altered by the presence of extra degrees of freedom in the modified gravity or DE model.

\subsection{Effective Dark Energy fluids in the $\{Q,\eta\}$ parametrization}\label{sec:DE-fluid-model}

As a particular realization, let us consider an effective DE fluid model which allows to cover both canonical scalar fields and K-essence models~\cite{Sapone-Kunz}. Depending on its internal properties, such a fluid can cluster over time and develop overdense/underdense regions, similarly to matter, and can also contribute to anisotropic stress if it has a non-zero viscosity. We can parametrize the fluid by three internal degrees of freedom; an EoS parameter $w$, a sound speed $c^2_s$ and a viscosity term $c^2_v$ damping density perturbations and responsible for anisotropic stress. Since the latter correspond to the off-diagonal components of the spatial energy momentum tensor $T^{i j}$, it represents the velocity of the $i$-th component of the pressure perturbation towards the $j$-th direction. Hence, $\sigma$ is sourced by the velocity perturbation $\theta$, modulated by a viscosity term $c_v$~\cite{Hu-viscosity}, and has the overall effect of damping the growth of perturbations when $c_v \neq 0$, see Ref.~\cite{DS-Finger3}.

Within this context, the clustering parameter $Q$ appearing in the modified Poisson \cref{eq:00-Q} can be regarded as describing the possible clustering of the DE fluid, i.e.
\begin{equation}
Q=\frac{\sum_i\rho_iD_i}{\rho_m D_m}=1+\frac{\rho_{\text{DE}} D_\text{DE}}{\rho_m D_m}\,,
\end{equation}
where $\rho_{\text{DE}}$ represents the DE density and $D_\text{DE}$ its corresponding density perturbation in the comoving gauge, as analogously defined for matter in \cref{comoving-gauge-density}. Following refs.~\cite{Sapone-Kunz,DS-Finger3}, for models in which the internal degrees of freedom are time-independent (but might still depend on the scale) $Q(a,k)$ can be written as
\begin{equation}\label{Q}
Q(a,k)=1+\frac{1-\Omega_m}{\Omega_m}(1+w)\frac{a^{-3w}}{1-3w+\frac{2k^2\hat{c}_s^2a}{3H_0^2\Omega_m}}\,,
\end{equation}
where the parameter $\hat{c}^2_s$ is regarded as an effective sound speed of the fluid given by
\begin{equation}\label{effective-c}
\hat{c}^2_s=c^2_s+\frac{8}{3}\frac{(c^2_s-w)}{(1+w)}c^2_{v}\,.
\end{equation}
From \cref{Q} it is clear that this kind of fluid is allowed to cluster only when $w\neq-1$, and this process is governed by $\hat{c}^2_s$. From \cref{effective-c} we note that the contribution due to $c^2_v$ can be enhanced with respect to that of $c^2_s$ by an order of magnitude for $w=-0.95$ and very small $c^2_s$, and then this is the regime where viscosity effects are expected to be more apparent. Furthermore, in ref.~\cite{DS-Finger3} it has been also shown that the anisotropy parameter for this DE fluid model is given by
\begin{equation}\label{eta}
\eta=-\frac{9}{2}H_0^2(1-\Omega_m)(1+w)\frac{a^{-(1+3w)}}{k^2Q}\left(1-\frac{c^2_s}{\hat{c}_s^2}\right)\,,
\end{equation} 
and vanishes identically if $c^2_v=0$. As shown in \cref{Q} and \cref{eta}, the magnitude of both $Q$ and $\eta$ grow with the scale factor and then in principle their effects should be more noticeable during late times. Finally, the growth index depends on the internal degrees of freedoms of this effective DE fluid as~\cite{Linder,DS-Finger3}
\begin{equation}\label{gindex}
\gamma(a,k)=\frac{3}{5-6w}\left(1-w-\frac{(1+\eta)Q-1}{1-\Omega_m(a)}\right)\,.
\end{equation}
The growth index~\cref{gindex} implies that the growth function $G$ \cref{definition-G} becomes scale dependent due to the presence of the sound horizon for the DE perturbations. In order to find analytical solution for the growth of the dark energy perturbations we required to assume a constant EoS $w$ and a constant sound speed $c_s$\footnote{the adiabatic sound speed $c_a$ is automatically set by the EoS. }. These assumptions are usually too strong and violated in dynamical dark energy models; however, we expect that the results still hold for models where these quantities vary slowly with time.

\section{The observed matter overdensity variable}\label{sec:matterdensity}

Probing models beyond $\Lambda$CDM through their impact on the clustering of dark matter represents the main goal of various large upcoming surveys such as Euclid~\cite{Euclid}, LSST~\cite{LSST} and DESI~\cite{Aghamousa:2016zmz}, but this can only be achieved by resorting to some form of luminous matter as tracer such as galaxies. It is now well-known that, besides bias effects, the observed distribution of galaxies includes several distortions arising from peculiar velocities and general relativistic effects. As shown in ref. \cite{Bonvin-Durrer}, assuming that the conservation equation for matter \cref{eq:v} holds, the \emph{observed} matter overdensity at redshift $z$ and direction in the sky $\hat{\textbf{n}}$ is given at the linear level by
\begin{equation}\label{dcontributions}
\Delta(z,\hat{\textbf{n}})=\sum_{i=1}^{7}\Delta_i\,,
\end{equation}
where the different contributions can be identified as follows:
\begin{align}
\Delta_D&=bD_s\,, \label{dmatter}\\
\Delta_{z}&=\frac{1}{\mathcal{H}}\partial_r(\textbf{V}\cdot\textbf{n})\,, \label{dRSD}\\
\Delta_V&=\left(\frac{\dot{\mathcal{H}}}{\mathcal{H}^2}+\frac{2}{r_s\mathcal{H}}\right)\textbf{V}\cdot\textbf{n}\,, \label{ddoppler}\\
\Delta_L&=-\frac{1}{r_s}\int_{0}^{r_s}dr\frac{r_s-r}{r}\Delta_{\Omega}({\Phi}+{\Psi})\,, \label{dlensing}\\
\Delta_{\text{lp}}&=\left(\frac{\dot{\mathcal{H}}}{\mathcal{H}^2}+\frac{2}{r_s\mathcal{H}}+1\right)\Psi-2\Phi+\frac{1}{\mathcal{H}}\dot{\Phi}\,, \label{dpotentials}\\
\Delta_{\text{std}}&=\frac{2}{r_s}\int_{0}^{r_s}dr(\Phi+\Psi)\,, \label{dshapiro}\\
\Delta_{\text{isw}}&=\left(\frac{\dot{\mathcal{H}}}{\mathcal{H}^2}+\frac{2}{r_s\mathcal{H}}\right)\int_{0}^{r_s}dr(\dot{\Phi}+\dot{\Psi})\,. \label{disw}
\end{align}
Here, $r_s=r_s(z)$ is the comoving distance to the source redshift given by \cref{eq:comoving-distance} and $\Delta_{\Omega}$ is the angular part of the Laplacian operator. The leading contribution to \cref{dcontributions} comes from \cref{dmatter}, where $\Delta_D$ represents the intrinsic fluctuations in the distribution of matter, related to the fact that baryons trace the underlying dark matter distribution (up to a linear bias factor $b(z)$), and all other terms represent distortions in the observed redshift and direction of incoming of photons, i.e. in the coordinate system in which we are making our observations; \cref{dRSD} corresponds to the RSD, which comes from the spatial gradient of peculiar velocities $\textbf{V}$ projected along the line-of-sight, while \cref{ddoppler} correspond to a Doppler effect which depends on the projection of $\textbf{V}$ along the line-of-sight. On the other hand, \cref{dlensing} correspond to weak gravitational lensing, and along the previous terms comprise the so-called $\textit{standard effects}$.

The remaining contributions, \cref{dpotentials}-\cref{disw}, are directly proportional to the Bardeen potentials $\Phi$ and $\Psi$ (i.e. not to their gradients), and represent the so-called \textit{relativistic effects}. These depend either locally on the source redshift, such as gravitational redshift and Sachs-Wolfe effect, or are integrated from source to observer along the line-of-sight. Following the convention adopted in ref.~\cite{Renk} the first class are identified collectively effects as the \textit{local potential terms} $\Delta_{\text{lp}}$, while the integrated terms correspond to the Shapiro time-delay effect $\Delta_{\text{std}}$ and the ISW effect $\Delta_{\text{isw}}$. The Shapiro time-delay reflects the fact that photons takes longer to travel through potential wells compared to flat space, which implies that we receive them slightly delayed and consequently more redshifted than if they travel only on the FLRW background. The ISW effect on the other hand considers fluctuations in the photons' energy as they travel from source to observer due to the time evolution of the potentials $\Phi$ and $\Psi$. It is worth noting that the relativistic effects are suppressed by a factor $(\mathcal{H}/k)^2$ with respect to the density contrast as can be seen by using the Poisson equation, \cref{eq:Poisson}, and by $\mathcal{H}/k$ with respect to peculiar velocities. Then, the magnitude of these effects might become substantial on horizon scales. We remark that \cref{dmatter}-\cref{disw} do not rely on the validity of Einstein equations but remain true for any model as long as the cosmic fluid is only coupled gravitationally, or equivalently, if matter follows geodesics~\cite{Bonvin-Durrer}. In the case of interactions \cref{eq:delta} and \cref{eq:v} can be violated and this result needs to be generalized~\cite{Duniya:2015nva,Duniya:2019}, while non-geodesic motion is strongly suppressed in most modified gravity models of interest due to the effect of screening mechanisms~\cite{Vainshtein:1972sx,Hui:2009kc,Ezquiaga17,Sakstein:2017xjx}. Then, the result \cref{dmatter}-\cref{disw} is compatible with the scope of this work (and with the DE fluid model introduced in section 2.2, in particular) and it is assumed as a valid linear approximation. We also ignore the effects on bias in \cref{dmatter} for the rest of this work.

\subsection{Angular power spectrum in the $\{Q,\eta\}$ parametrization}
\label{sec:angular-ps}

We next focus our attention on writing down an expression for the angular power spectrum $C_\ell$ for the full observed density perturbation variable \cref{dcontributions} considering the $\{Q,\eta\}$ parametrization. In particular, this allows us to study the effective DE fluid model discussed in \cref{sec:DE-fluid-model}, but can be also applied to any model fitting within this framework. We start by expanding the matter overdensity in terms of spherical harmonics as
\begin{equation}
\Delta(z,\hat{\textbf{n}})=\sum_{\ell m}a_{\ell m}(z)Y_{\ell {m}}(\hat{\textbf{n}})\,,
\end{equation}
where
\begin{equation}\label{eq:alm}
a_{\ell m}(z)=\int d\Omega_{\hat{\textbf{n}}}Y^*_{\ell {m}}(\hat{\textbf{n}})\Delta(z,\hat{\textbf{n}})\,.
\end{equation}
The angular power spectrum is then given by $C_{\ell}(z)=\langle\vert a_{\ell m}(z)\vert^2\rangle$, where the brackets indicate ensemble average as encoded in the nature of the primordial power spectrum \cref{eq:primordialPS}. The $C_{\ell}$ has the advantage of being optimally adapted to our coordinate system in which we perform the measurements as it exploits the statistical isotropy upon which the FLRW metric is constructed. In order to calculate \cref{eq:alm} we consider the fact that according to \cref{dcontributions}-\cref{disw} the observed overdensity consists in the linear combination of terms whose $k$-dependences are given by either a perturbation variable evaluated at the source position $r_s={\tau}_o-{\tau}_s$ (i.e. at a fixed redshift) or by an integral of a perturbation variable over the unperturbed photon trajectory (assuming the Born approximation). Using \cref{Td}-\cref{Tphi}, but without using the field equations yet, the full expression for the angular power spectrum is then
\begin{equation}\label{eq:full-cl}
C_{\ell}(z)=\frac{2A_{s}}{\pi}\int_{0}^\infty\frac{dk}{k}(k{\tau}_0)^{n_s-1}|\Delta_{\ell}(z,k)|^2\,,
\end{equation}
where the different contributions to $\Delta_{\ell}(z,k)=\sum_i\Delta^i_{\ell}$ are 
\begin{align}
\Delta^D_{\ell}&=j_{\ell}T_D\,, \label{dlmatter}\\
\Delta^{z}_{\ell}&=\frac{k}{\mathcal{H}}j_{\ell}''T_V\,, \label{dlRSD}\\
\Delta^L_{\ell}&=\frac{1}{r_s}\int_{0}^{r_s}dr j_{\ell}(kr)\frac{r_s-r}{r}\ell(\ell+1)(T_\Psi+T_\Phi)\,, \label{dllensing}\\
\Delta^V_{\ell}&=j_{\ell}'\left(\frac{\dot{\mathcal{H}}}{\mathcal{H}^2}+\frac{2}{r_s\mathcal{H}}\right)T_V\,, \label{dldoppler}\\
\Delta^{\text{lp}}_{\ell}&=j_{\ell}\left[\left(\frac{\dot{\mathcal{H}}}{\mathcal{H}^2}+\frac{2}{r_s\mathcal{H}}+1\right)T_\Psi+T_\Phi+\frac{1}{\mathcal{H}}\dot{T}_\Phi\right]\,, \label{dlpotentials}\\
\Delta^{\text{std}}_{\ell}&=\frac{2}{r_s}\int_{0}^{r_s}dr j_{\ell}(kr)(T_\Psi+T_\Phi)\,, \label{dlshapiro}\\
\Delta^{\text{isw}}_{\ell}&=\left(\frac{\dot{\mathcal{H}}}{\mathcal{H}^2}+\frac{2}{r_s\mathcal{H}}\right)\int_{0}^{r_s}dr j_{\ell}(kr)(\dot{T}_\Phi+\dot{T}_\Psi)\,. \label{dlisw}
\end{align}
Here, $j_\ell(x)$ represents a spherical Bessel function which appear due to the Rayleigh formula for plane waves and $j'_{\ell}(x)\equiv dj_{\ell}(x)/dx$. Then, the total power spectrum $C_{\ell}$ is given by
\begin{equation}
C_{\ell}=\sum_{ij}C^{ij}_{\ell}\,,
\end{equation}
where each contribution (either autocorrelation or cross-correlation) can be systematically computed as
\begin{equation}\label{eq:cij}
C^{ij}_\ell=\frac{2A_{s}}{\pi}\int_{0}^\infty\frac{dk}{k}(k{\tau}_0)^{n_s-1}\Delta^i_\ell\Delta^j_\ell\,.
\end{equation}
The previous equation, along with \cref{dlmatter}-\cref{dlisw} give the generic structure of the power spectrum regardless of the relation between the metric and matter degrees of freedom as we have not imposed the field equations yet.  We note that the $C_{\ell}$'s depend explicitly on the background cosmology through the conformal Hubble parameter $\mathcal{H}$, its derivative $\dot{\mathcal{H}}$ and the comoving distance $r_s$, which are sensitive to the EoS parameter of the DE fluid, while at the perturbation level they depend on the growth factor $G$ as well as on the relation among the various transfer functions $\{T_D, T_V, T_\Psi, T_\Phi\}$. If we now adopt the $\{Q,\eta\}$ scheme, all these transfer functions can be completely parametrized in terms of $T(k)$ and $G$ through \cref{TD-eq}-\cref{Tphi-eq}. The final expressions for the leading terms contributing to \cref{eq:cij} calculated under this scheme are included in \cref{sec:appendix}. Given a particular form for $\{Q,\eta\}$ (i.e. a particular DE or modified gravity model) these allow to carry out a direct integration for a given cosmology.

\section{Results}\label{sec:results}

In this section we investigate the observed angular power spectrum of the effective DE fluid model in the $\{Q,\eta\}$ parametrization discussed in \cref{sec:DE-fluid-model}, considering both standard and relativistic effects. For the DE fluid model we consider a fixed EoS parameter $w=-0.95$ and restrict the effective sound speed to the range $10^{-6}\leq \hat{c}_s^2\leq 1$. The results are compared with respect to a fiducial $\Lambda$CDM cosmology using the Planck 2018 best-fit parameters~\cite{Planck18}. 

Notice that, for this model, the $z=0$ transfer function $T(k)$ is the only numerical input entering in the $C_\ell$'s (see \cref{eq:ClDD}-\cref{eq:CliswD} for more details). We compute this using a modified version of {\sc camb} for the DE fluid model, and with the standard {\sc camb} code~\cite{CAMB} for $\Lambda$CDM (no Halofit used). In addition, since in this model the DE behavior departures from $\Lambda$ at late times, we normalize the growth function~\cref{definition-G} to be unity well into the matter domination era in all cases, i.e. $G(a_0=10^{-3},k)=1$, as the DE fluid mimics $\Lambda$ during that regime. In order to quantify deviations from $\Lambda$CDM, we define the relative power spectrum of each effect contributing to \cref{eq:cij} as
\begin{equation}\label{delta-cij}
\Delta C^i_\ell=1-\frac{C^{i}_\ell(\text{DE})}{C^{i}_\ell(\Lambda)}\,,
\end{equation}
where $C^{i}_\ell(\text{DE})$ and $C^{i}_\ell(\Lambda)$ are the power spectra for a given effect calculated under the DE model and $\Lambda$CDM, respectively. We stress that, as a way to disentangle the origin of deviations from $\Lambda$CDM, in \cref{delta-cij} we only consider autocorrelations. Complementarily, to keep track of the relative amplitude of each effect it is also useful to quantify the relative power spectrum with respect to the total signal predicted by $\Lambda$CDM. This is given by
\begin{equation}\label{delta-cij-tot}
\Delta C^{i-\text{tot}}_\ell=\frac{C^{i}_\ell(\text{DE})-C^{i}_\ell(\Lambda)}{C^{\text{tot}}_\ell(\Lambda)}\,,
\end{equation}
where each term in the numerator represents the total contribution coming from a given effect, i.e. including both its autocorrelation and cross-correlation with density, and $C^{\text{tot}}_\ell(\Lambda)$ correspond to the $\Lambda$CDM  power spectrum with all contributions to \cref{eq:cij}.

\subsection{The perfect Dark Energy fluid case}\label{sec:perfect-DE-fluid}

\begin{figure}[tbp]
	\centering 
	\includegraphics[width=1\textwidth]{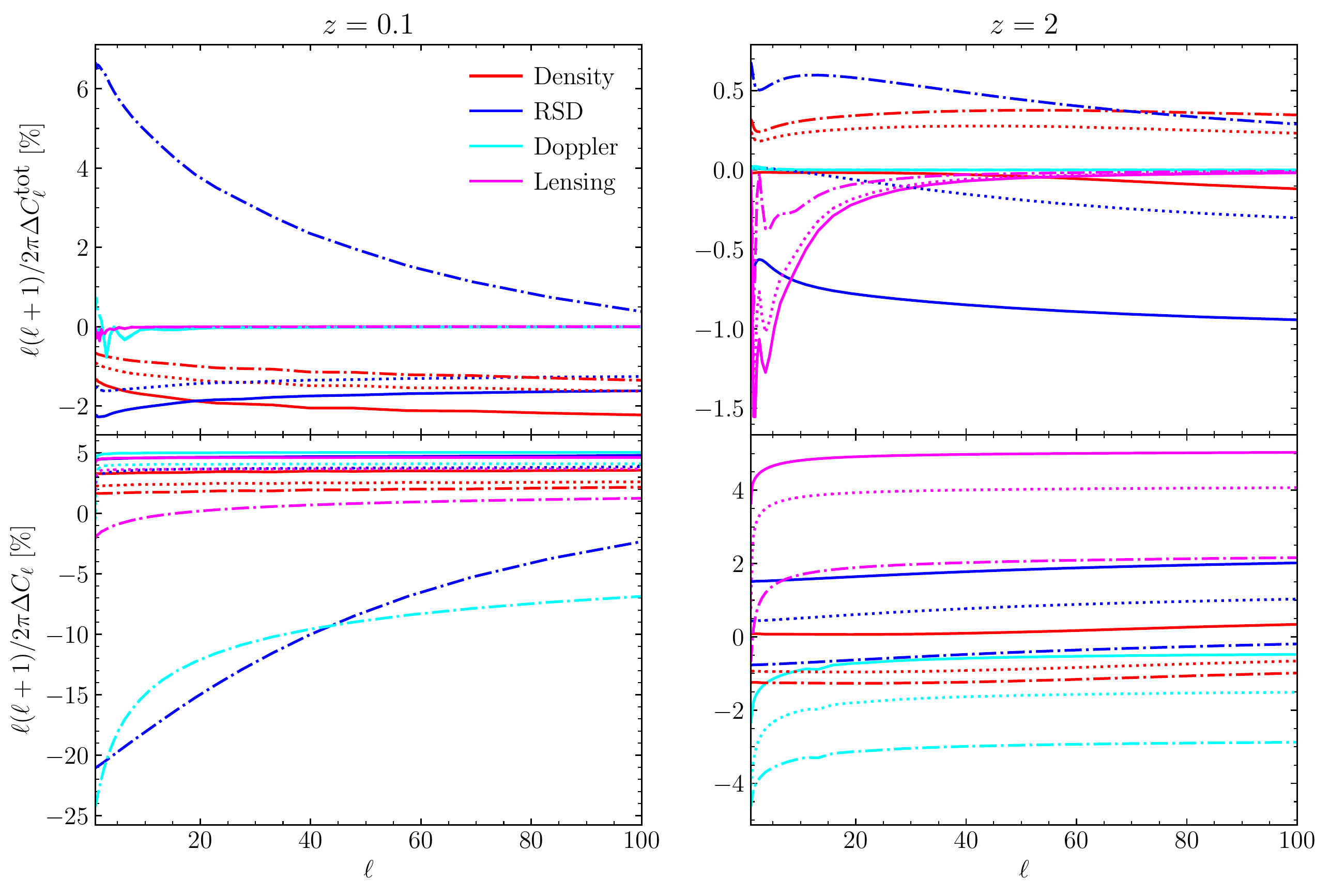}
	\caption{\label{fig:standard-effects} Deviations in the Standard effects for $z=0.1$ (left column) and $z=2$ (right column). Pure effects (autocorrelations only) are shown in the bottom row, while the top row shows deviations relative to the total $\Lambda$CDM power spectrum when cross-correlation with density is included in each effect. The different line styles represent the cases $c^2_s=1$ (solid), $c^2_s=10^{-2}$ (dotted) and $c^2_s=10^{-6}$ (dot-dashed).}
\end{figure}

Let us start by investigating the angular power spectrum of an effective DE fluid in absence of anisotropic stress, i.e. with $\eta=c^2_v=0$. For this, we consider the cases $c^2_s=10^{-6}$, $c^2_s=10^{-2}$ and $c^2_s=1$, the latter corresponding to a completely smooth DE component similar to the cosmological constant. \cref{fig:standard-effects} (top row) shows the total relative deviations of the standard effects obtained with \cref{delta-cij-tot}. We find that, at low redshift, the intrinsic matter density fluctuations $D$ (red) and RSD (blue) dominates the amplitude of deviations from the full $\Lambda$CDM signal in all cases. For the case $c^2_s=10^{-6}$ the latter effect is particularly important as we move towards large scales. At $z=2$, the RSD deviations exhibit the largest amplitude among the standard effects except at large scales, where this is surpassed by the lensing signal (magenta). On the other hand, the relative weight of the Doppler effect deviations (cyan) remain small at both redshifts.

In addition, \cref{fig:standard-effects} (bottom row) shows the relative deviations in the standard effects calculated with \cref{delta-cij}, i.e. comparing autocorrelations of each individual effect against their $\Lambda$CDM counterparts. To understand the impact of the DE fluid model on the different $C_\ell$'s we note that the matter power spectrum $P(k)$ (and hence the transfer function $T(k)$) receives both background and perturbative contributions through the model parameters $w\geq-1$ and $\hat{c}^2_s\leq1$, respectively. Firstly, for $w=-0.95$ the DE domination era begins earlier than in $\Lambda$CDM and then there is an extra suppression of structure formation, which translates into lower amplitudes of $P(k)$ and $T(k)$ at the present day. This leads to a positive $\Delta C_\ell$ according to our sign convention in \cref{delta-cij}. Secondly, as $\hat{c}^2_s$ departs from unity the DE perturbations can start to grow outside the sound horizon, leading to an enhancement of $T(k)$ that competes against the suppression due to $w>-1$. Then, in~\cref{fig:standard-effects} we find that matter density fluctuations at $z=0.1$ are suppressed the most for the smooth case ${c}^2_s=1$, while for lower sound speeds the extra clustering compensates the effect of $w$ and the amplitudes become closer to $\Lambda$CDM. However, at $z=2$ the relative difference between the expansion histories described by $w=-0.95$ and $w=-1$ is reduced and the contribution from the clustering parameter $Q$ dominates, leading to $\Delta C_\ell>0$ and larger deviations in the lowest sound speed scenarios. The fact that the low sound speed models become closer to $\Lambda$CDM due to the DE perturbations is consistent with the results discussed in ref.~\cite{Duniya:2013eta}, although considering a different normalization of the matter power spectra which removes the effects of the background expansion at $z=0$ but keeps the effect of the quintessence clustering on large scales. % Density.

From \cref{fig:standard-effects} we also find that at $z=0.1$ deviations in RSD and Doppler effect are greatly enhanced with respect to $\Lambda$CDM for $\cs2=10^{-6}$, which reflects the impact of the DE fluid clustering on peculiar velocities as well as on its divergence. Both effects probe the growth factor $G$ and the clustering parameter $Q$ through the combinations $GQ>0$ and $\partial(GQ)/\partial{a}<0$, which then compete against each other, see \cref{zz}-\cref{VD}. At low redshift the rate of change $\partial(GQ)/\partial{a}$ has its largest impact (since $G=1=\text{const.}$ as we reach matter domination era) while at high redshift the term $GQ$ eventually dominates and deviations decay. Notice that the amplitude of relative lensing deviations do not decay towards higher $z$ as strongly as the previous effects. Since this effect integrates the projections of $\Phi$ and $\Psi$ along the line of sight, increasing $z$ allows this to probe the DE model over a larger portion of the universe. % RSD, Doppler and lensing.

\begin{figure}[tbp]
	\centering 
	\includegraphics[width=1\textwidth]{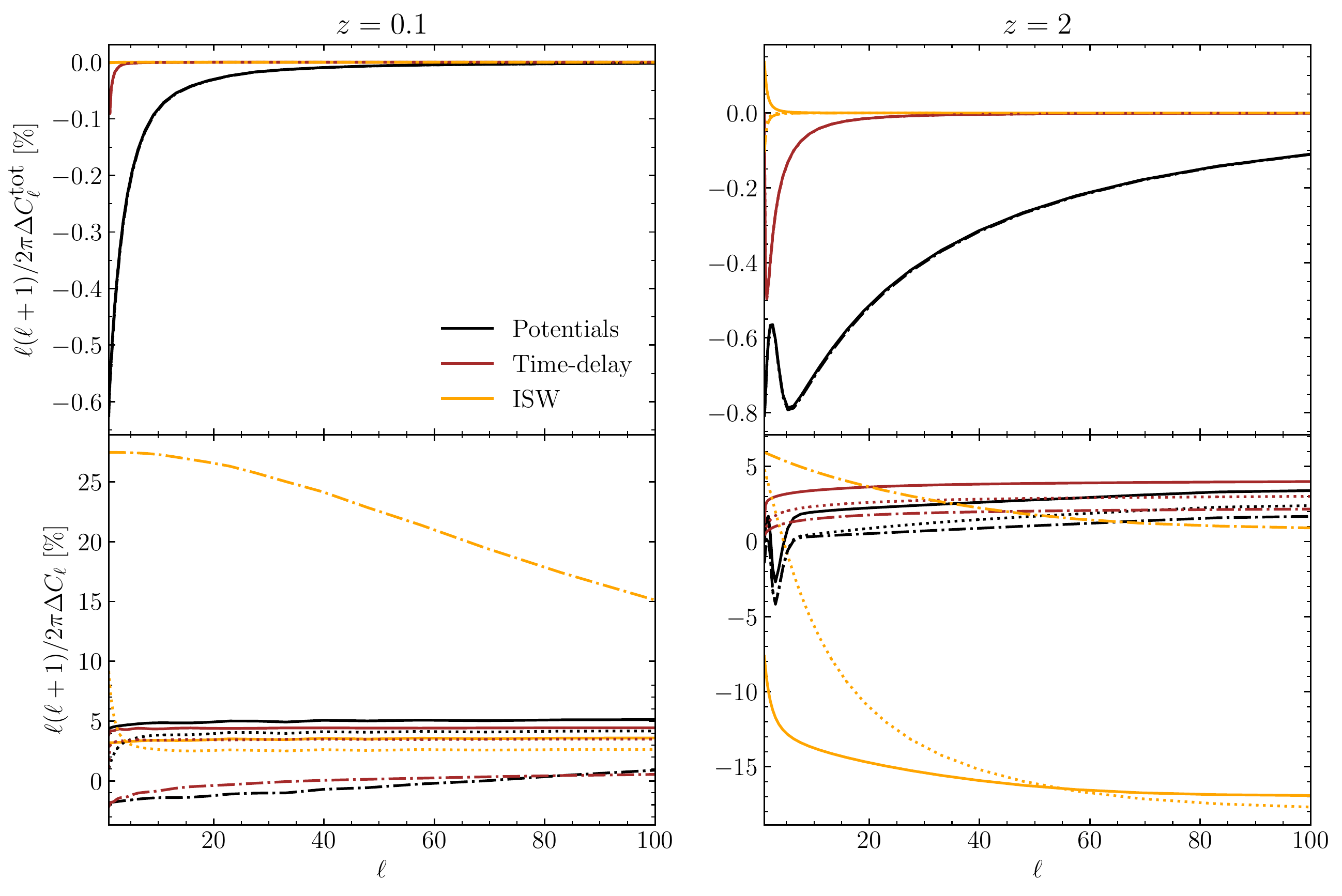}
	\caption{\label{fig:relativistic-effects} Deviations in the Relativistic effects for $z=0.1$ (left column) and $z=2$ (right column). Pure effects (autocorrelations only) are shown in the bottom row, while the top row shows deviations relative to the total $\Lambda$CDM power spectrum when cross-correlation with density is included in each effect. The different line styles represent the cases $c^2_s=1$ (solid), $c^2_s=10^{-2}$ (dotted) and $c^2_s=10^{-6}$ (dot-dashed).}
\end{figure}

On a similar fashion, \cref{fig:relativistic-effects} (top row) shows the total deviations in the relativistic effects relative to the full $\Lambda$CDM power spectrum. We find that at both redshifts the signal is dominated by the local potential terms (black) at very large scales but decays rapidly.  The other two relativistic effects, i.e. Shapiro time-delay (brown) and ISW (orange) remain subdominant relative to the full signal in all cases. From the autocorrelation-only terms shown in~\cref{fig:relativistic-effects} (bottom row) we find that, at low redshift, modifications to the Shapiro time-delay and the local potential terms have a similar behavior since for nearby sources both effects carry roughly the same information. Remarkably, for $c^2_s=10^{-6}$ deviations in the ISW effect at $z=0.1$ can be considerably larger than for the other two relativistic effects despite the fact that we are integrating over a narrow redshift range. This is due to the fact that in this DE model the evolution of the metric potentials $\Psi$ and $\Phi$ is quickly deviating from the $\Lambda$CDM behavior at late times due to the clustering of the DE fluid, and the power spectrum is more sensitive to the rate of change $\dot{\Phi}$ than to $\Phi$ itself. In addition, we find that at $z=2$ deviations in the ISW effect are stronger for the higher sound speed scenarios, where the DE component is smoother, and stills dominate over the other two effects. While at $z=0.1$ the background coefficient $({\dot{\mathcal{H}}}/{\mathcal{H}^2}+{2}/{r_s\mathcal{H}})^2$ appearing in \cref{isw} for the ISW effect deviates less than $1\%$ with respect to the $\Lambda$CDM expansion history, this can enhance this effect by $\sim15\%$ at $z=2$ which shifts the ISW curves shown in~\cref{fig:relativistic-effects} toward negatives values.

The high sensitivity of the ISW effect at both low and high redshift can be understood by the fact that it cumulatively probes the combination $\partial(GQ)/\partial{a}$, whose deviations with respect to $\Lambda$CDM at late times can be larger than those coming from $GQ$ itself. Notice that, as we discussed previously, the combination $\partial(GQ)/\partial{a}$ appears in the RSD and Doppler effects, which show the largest deviation among the standard effects, but in the ISW effect this is boosted by the integration from source to observer. In addition, in such standard effects $\partial(GQ)/\partial{a}$ has to compete against the combination $GQ$ (which has opposite sign) while the ISW effect probes the former in isolation. In fact, it has been widely remarked that the rate of change $\partial{G}/\partial{a}$ appearing naturally through the ISW effect is an excellent discriminator of DE models \cite{Huterer-Growth}, as well as the `clustering rate' $\partial{Q}/\partial{a}$~\cite{Sapone-Kunz,DS-Finger3}. Since in $\Delta C_\ell$ we compare the DE fluid model relative to $\Lambda$CDM, according to \cref{isw} the overall enhancement of the ISW effect at the perturbative level is governed by the so-called magnification parameter~\cite{Sapone-Kunz}
\begin{equation}\label{amp}
\mathcal{A}\equiv\frac{\partial(G(a,k)Q(a,k))/\partial a}{\partial G_\Lambda(a)/\partial a}=Q\frac{\partial G/\partial a}{\partial G_\Lambda/\partial a}+\frac{G}{\partial G_\Lambda/\partial a}\frac{\partial Q}{\partial a}\,.
\end{equation}
Then, when testing a smooth DE model only the first term in \cref{amp} contributes (since $Q=\text{const.}$), while for low sound speeds such as in the $\cs2=10^{-6}$ case the clustering rate $\partial{Q}/\partial{a}$ provides and extra contribution to \cref{amp} and greatly enhances the ISW signal relative to $\Lambda$CDM. Notice that this behavior is opposite to that of $T(k)$, which becomes closer to $\Lambda$CDM for low sound speeds and can break the hierarchy in the ISW curves shown in \cref{fig:relativistic-effects} since $\Delta C^{\text{isw}}_\ell\propto (T_\text{DE}/T_\Lambda\times\mathcal{A})^2$, where $T_\text{DE}$ and $T_\Lambda$ are the transfer functions obtained using the DE model and $\Lambda$CDM, respectively. However, $\mathcal{A}$ is highly sensitive to $\cs2$ through the clustering rate and dominates the overall enhancement of the ISW effect. A summary of the deviations in standard and relativistic effects in the perfect DE fluid model at large and small scales is included in \cref{table:summary-deviations}.

\begin{table}[h]
	\begin{center}
		\begin{tabular}{ l | c | c | c | c | c | c }
			&\multicolumn{2}{c}{$\cs2=1$} &\multicolumn{2}{|c}{$\cs2=10^{-2}$}&\multicolumn{2}{|c}{$\cs2=10^{-6}$}\\
			\hline
			\hline
			\text{effect}& $z=0.1$ & $z=2$ & $z=0.1$ & $z=2$ & $z=0.1$ & $z=2$\\
			\hline
			\hline
			$\text{Density}$  &$3$ $(3)$  &$0$ $(0)$ &$2$ $(2)$ &$-1$ $(-1)$  &$2$ $(2)$ &$-1$ $(-1)$\\ 
			$\text{RSD}$      &$4$ $(3)$  &$2$ $(2)$   &$3$ $(3)$ &$0$ (1)   &$-20$ $(-5)$ &$-1$ $(0)$\\ 
			$\text{Doppler}$  &$5$ $(5)$  &$-2$ $(3)$  &$4$ $(4)$ &$-2$ $(-2)$  &$-20$ $(-7)$ &$-4$ $(-3)$\\
			$\text{Lensing}$  &$4$ $(4)$ &$5 (5)$  &$4$ $(3)$ &$4$ $(4)$   &$-1$ $(0)$  &$2$ $(2)$\\
			\hline
			$\text{Local P.}$ &$5$ $(5)$  &$-1$ $(3)$  &$3$ $(3)$ &$-2$ $(2)$  &$-2$ $(0)$ &$-2$ $(2)$\\ 
			$\text{Shapiro}$  &$4$ $(4)$  &$3$ $(4)$   &$3$ $(3)$ &$2$ $(3)$   &$-1$ $(0)$ &$1$ $(2)$\\ 
			$\text{ISW}$      &$3$ $(3)$  &$-13$ $(17)$&$3$ $(3)$ &$-2$ $(-17)$ &$27$ $(18)$  &$5$ $(1)$\\
			\hline
		\end{tabular}
		\caption{Summary of deviations in standard and relativistic effects in the perfect DE fluid model with respect to $\Lambda$CDM. The values (rough percentages) outside the parenthesis correspond to $\ell=5$, while values in parenthesis correspond to deviations at $\ell=80$.}\label{table:summary-deviations}
	\end{center}
\end{table}

\subsection{The imprints of viscosity}\label{sec:viscosity}

Let us now go beyond the perfect DE fluid model discussed in \cref{sec:perfect-DE-fluid} and study the imprints of a potential fluid viscosity in standard and relativistic effects. In our model, this implies the presence of an additional sound speed parameter $c^2_v$ which give rise to a non-vanishing anisotropic stress, so that $\Phi\neq\Psi$ in this case. Moreover, there is an effective sound speed $\hat{c}^2_s$ given by a combination of $c^2_s$ and $c^2_v$, see~\cref{effective-c}, and thus this new parameter might introduce some degeneracies in our observables. Here we consider two possible scenarios given by the combinations $(c^2_s,c^2_v)=(0,2\times 10^{-6})$ and $(c^2_s,c^2_v)=(10^{-6},2\times10^{-4})$, corresponding to the effective sound speeds $\hat{c}^2_s\approx10^{-4}$ and $\hat{c}^2_s\approx10^{-2}$, respectively. 

%\begin{figure}[tbp]
%	\centering 
%	\includegraphics[width=.49\textwidth]{fig/cv_auto_std_effects_z01_subplot_v2}
%	\hfill
%	\includegraphics[width=.49\textwidth]{fig/cv_auto_std_effects_z2_subplot_v2}
%	
%	\includegraphics[width=.49\textwidth]{fig/v2/cv_auto_rel_effects_z01_subplot_v2}
%	\hfill
%	\includegraphics[width=.49\textwidth]{fig/v2/cv_auto_rel_effects_z2_subplot_v2}
%	\caption{\label{fig:all-effects-cv} Deviations in the Standard effects (top row) and Relativistic effects (bottom row) at two different source redshifts considering autocorrelations only. Left column: $z=0.1$, right column $z=2$. The line styles represent the cases $c^2_s=10^{-2}$ (solid), $(c^2_s,c^2_v)=(10^{-6},2\cdot10^{-4})$ (doted), $c^2_s=10^{-4}$ (dashed) and $(c^2_s,c^2_v)=(0,2\cdot10^{-6})$ (dot-dashed). Notice that in these we have used logarithmic scale in the horizontal axis for a better comparison of the models.}
%\end{figure}

\begin{figure}[tbp]
	\centering 
	\includegraphics[width=1\textwidth]{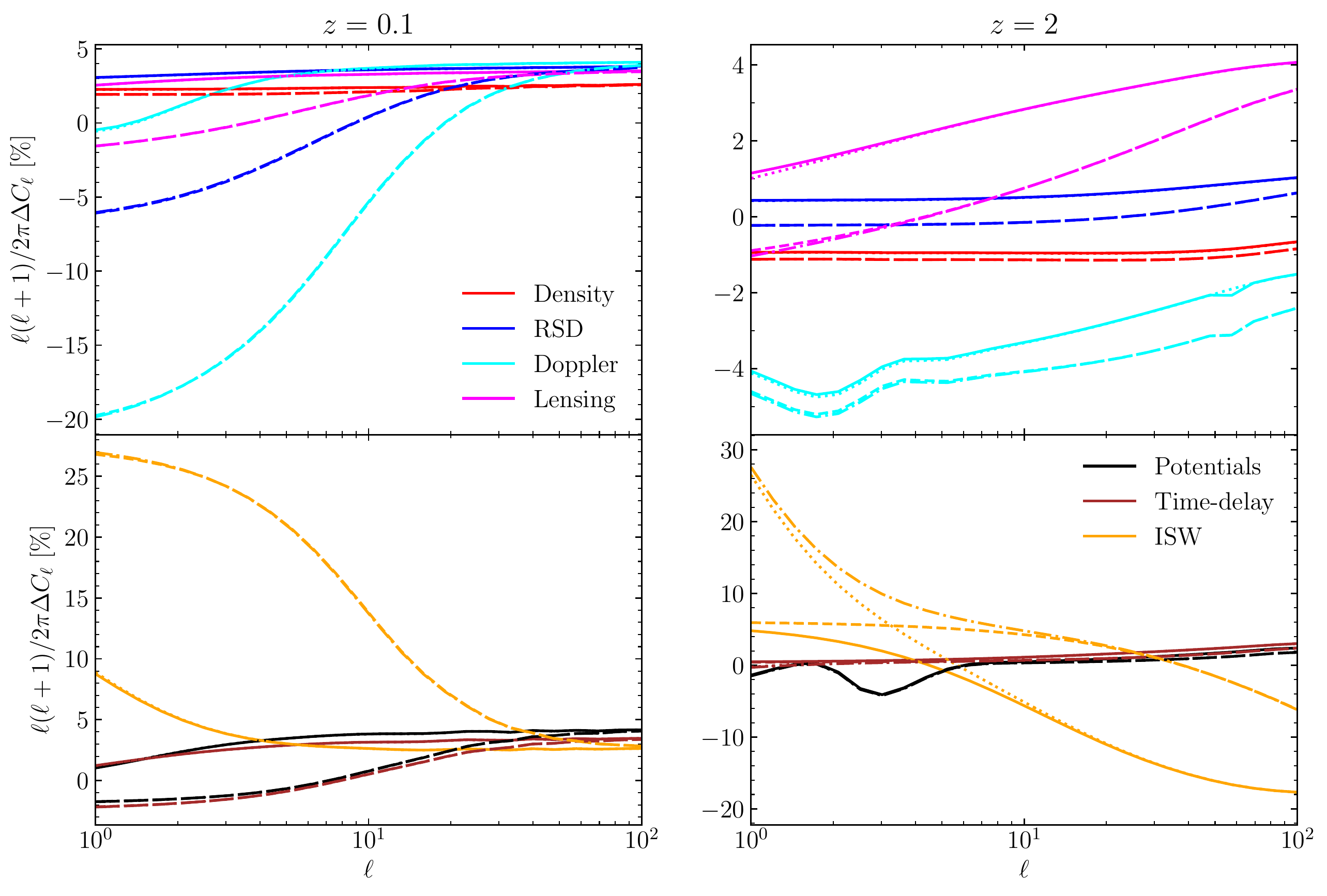}
	\caption{\label{fig:all-effects-cv} Deviations in the Standard effects (top row) and Relativistic effects (bottom row) at two different source redshifts considering autocorrelations only. Left column: $z=0.1$, right column $z=2$. The line styles represent the cases $c^2_s=10^{-2}$ (solid), $(c^2_s,c^2_v)=(10^{-6},2\cdot10^{-4})$ (doted), $c^2_s=10^{-4}$ (dashed) and $(c^2_s,c^2_v)=(0,2\cdot10^{-6})$ (dot-dashed). Notice that in these we have used logarithmic scale in the horizontal axis for a better comparison of the models.}
\end{figure}

\cref{fig:all-effects-cv} (top row) shows the power spectra of the standard effects (relative to $\Lambda$CDM) for the models with viscosity as well as for their corresponding perfect fluids counterparts which have the same effective sound speed but $c^2_v=\eta=0$ (notice that in these we have used logarithmic scale in the horizontal axis for a better comparison). According to \cref{Q} the clustering parameter $Q$ does not depend on the intrinsic sound speeds but on $\hat{c}^2_s$, so that it is degenerated with respect to $c^2_s$ and $c^2_v$. However, \cref{gindex} shows that the growth index $\gamma$ (which appears in the growth rate $G$) does depend directly on $c^2_v$ through $\eta$. From \cref{fig:all-effects-cv} we find that, nonetheless, the standard effects cannot distinguish between viscosity and non-viscosity scenarios as there is negligible difference between power spectra for the same effective sound speeds at both redshifts. 

On the other hand, from the relativistic effects shown in~\cref{fig:all-effects-cv} (bottom row) we find that the ISW effect is able to reveal the presence of viscosity at high redshift and very large scales. This arise from its advantage of probing the rate of change of the anisotropic stress $\partial\eta/\partial a$ in a cumulative way rather than $\eta$ itself, which is analogous to the enhancement due to $\partial{Q}/\partial{a}$ discussed in the previous subsection so that the ISW effect is greatly boosted compared to $\Lambda$CDM. Then, the contribution of $\eta$, despite being small, is able to introduce sizeable deviations in the ISW effect from the $\hat{c}^2_s=c^2_s$ counterpart. As \cref{eta} shows, since $\eta\propto k^{-2}$ this is mostly important at very large scales, which is consistent with these and previous results~\cite{DS-Finger3}.

At this point it is important to remark that in DE models, the propagation of gravitational waves can be modified by the presence of the extra degrees of freedom. In particular, it has been shown that in scalar-tensor theories the anisotropic stress generally modifies the propagation speed of tensor modes, $c_T$~\cite{Amendola:2017orw}. This means that currently the magnitude of $\eta$ is strongly constrained by the tiny deviation in $c_T$ from the speed of light observed in the events GW170817 and GRB170817A~\cite{Ligo-2017}. However, scalar-tensor theories in which the anisotropic stress is scale-independent, as well as theories beyond the Horndeski class can avoid this constraint~\cite{Nersisyan:2018auj}. In the case of the particular model considered in this subsection, the constraint $|c_T-1|\leq 5\cdot10^{-16}$ puts a bound on the maximum value of the viscosity parameter of $c^2_v\lesssim\mathcal{O}(10^{-12})$, and then these results should only be regarded as illustrating the type of impact of viscosity in the various relativistic effects that could arise in DE models. Finally, we would like to highlight that the choice of the anisotropic stress is a difficult task as it depends on the microphysics of the fluid, if this is real, or on the modification of the geometry of spacetime, if the fluid is an effective one, and then the $k$-dependence of $\eta$ is due to the particular choice of anisotropic stress equation of motion. Moreover, in the former case the anisotropic stress of the matter sources does not modify the homogeneous part of the wave equation satisfied by the tensor modes~\cite{Saltas:2014dha}.

\section{Conclusions}

In this paper, we have studied the observed angular power spectrum of effective K-essence and quintessence DE fluids considering standard and relativistic effects. For this purpose, we adopted the phenomenological approach that introduces two functions $\{Q,\eta\}$ at the level of linear perturbations and allow to conveniently parametrize the modified clustering (or effective gravitational constant) and anisotropic stress appearing in models beyond $\Lambda$CDM. Under this particular framework, we have derived expressions for the thirteen most dominant contributions in the observed angular power spectrum of galaxies. We found that, overall, deviations relative to $\Lambda$CDM are stronger at low redshift since the behavior of the DE fluid can mimic the cosmological constant during matter domination era but departs during the late time universe. In particular, at $z=0.1$ the matter density fluctuations are suppressed by up to $\sim3\%$ for the quintessence-like case, while RSD and Doppler effect can be enhanced by $\sim15\%$ at large scales for the lowest sound speed scenario. On the other hand, at $z=2$ integrated effects become more important since lensing deviation can reach $\sim5\%$, whereas the ISW effect can deviate up to $\sim17\%$ due to its capacity to probe cumulatively not only $G$ itself but also its rate of change, which agrees with previous studies \cite{Sapone-Kunz,Huterer-Growth}. 

Furthermore, when considering an imperfect DE fluid scenario, we find that all effects are insensitive to the presence of anisotropic stress at low redshift and only the ISW effect can detect this feature at $z=2$ and very large scales, which reaffirms its power for testing and constraining DE models using future galaxy surveys. However, the amplitude of this effect is the weakest of the full observed matter power spectrum, and then it needs to be carefully extracted from the full signal. The fact that the imprints of relativistic effects appear at large scales poses a challenge for their detectability due to cosmic variance, but techniques such as multitracers might allow to overcome this issue~\cite{Seljak:2009,Yoo-Seljak:2012}.

Finally, we remark that the angular power spectrum expressions included in~\cref{sec:appendix} are not restricted to the effective DE fluid model investigated in this work but they can be applied to a broad class of DE and modified gravity models that can be encompassed in this effective $\{Q,\eta\}$ framework (and in which geodesic motion is not violated), such as $f(R)$ gravity and the DGP model, among others. For this, the explicit forms for the functions $\{Q,\eta\}$ are needed and come from matching \cref{eq:00-Q} and \cref{eq:ij-eta} with the linear field equations of the theory, as well as the growth factor $G$ (or growth index). For well-studied models such as the previously mentioned these can be found in the literature~\cite{Linder:2005in,Narikawa:2009ux,Basilakos:2012uu}, while the transfer functions at $z=0$ can be calculated from linear codes for modified gravity such as {\sc mgcamb}~\cite{MGCAMB}.

\acknowledgments
The authors are grateful to Guillermo Blanc, Gonzalo Palma and Nelson Zamorano for useful comments and discussion. We also thank the anonymous referee for their constructive feedback and comments. DS and CB-H acknowledge support from the Chilean National Commission for Scientific and Technological Research (CONICYT) through FONDECYT grant 11140496. CB-H acknowledges support through CONICYT/Becas-Chile 72180214.

\appendix
\section{Angular power spectrum expressions}\label{sec:appendix}

In this appendix we include the general expressions for the 13 most dominant contributions to the total angular power spectrum~\cref{eq:full-cl} in terms of the $\{Q,\eta\}$, and correspond to the autocorrelations of the different effects and their cross-correlations with density perturbations. The transfer function at redshift zero $T(k)$ is computed numerically with a modified version of {\sc camb}\footnote{we are explicitly including the factor $Q(a,k)$ in these expressions so that $T(k)$ correspond to the actual transfer function at $z=0$ from {\sc camb} codes.}. Notice that we have omitted the $(a,k)$ dependence in most terms to avoid cluttered notation. We also denote $j'_{\ell}(x)=dj_{\ell}(x)/dx$ and $\nu=\ell+1/2$ for the Limber-approximated terms \cite{Limber}.

\subsection{Standard Effects}
\paragraph{Density:} The dominant term in the observed matter power spectrum is the autocorrelation of intrinsic density fluctuations, which is given by
	\begin{align}
	C_{\ell}^{DD}=\frac{8A_{s}{\tau}^{n_s-1}_0a^2}{9\pi H^4_0\Omega^2_m}\int_{0}^{\infty}{dk}k^{2+n_s}\frac{j_{\ell}^2(kr_s)G^2T^2(k)}{(1+\eta)^2}\,.\label{eq:ClDD}
	\end{align}
\paragraph{Redshift-space distortion:}
	
\begin{subequations}
The next-to-leading contribution to the observed matter power spectrum is the autocorrelation of RSD, which is given by
		\begin{align}
		C_{\ell}^{zz}=&\frac{8A_{s}{\tau}^{n_s-1}_0a^2}{9\pi H^4_0\Omega^2_m}\times\label{zz}\\
		&\int_{0}^{\infty}{dk}{k^{2+n_s}}\left[\left(1-\frac{a}{(1+\eta)^2}\frac{\partial\eta}{\partial a}\right)GQ+\frac{a}{1+\eta}\frac{\partial(GQ)}{\partial a}\right]^2(j''_{\ell}(kr_s))^2T^2(k)\,.\nonumber
		\end{align}
The cross-correlation between density perturbations and RSD is also particularly important since it scales in the same way with $(k/\mathcal{H})$ as the autocorrelations of such effects. This is given by
		\begin{align}
		C_{\ell}^{Dz}=&-\frac{8A_{s}{\tau}^{n_s-1}_0a^2}{9\pi H^4_0\Omega^2_m}\int_{0}^{\infty}{dk}{k^{2+n_s}}\left[\left(1-\frac{a}{(1+\eta)^2}\frac{\partial\eta}{\partial a}\right)GQ+\frac{a}{1+\eta}\frac{\partial(GQ)}{\partial a}\right]\nonumber\\
		&\times\frac{j_{\ell}(kr_s)j''_{\ell}(kr_s)GT^2(k)}{(1+\eta)}\,.
		\end{align}
\end{subequations}
	
\paragraph{Doppler effect:} The autocorrelation of the Doppler effect is
\begin{subequations}
		\begin{align}
		C_{\ell}^{VV}&=\frac{8A_{s}{\tau}^{n_s-1}_0a^2}{9\pi H^4_0\Omega_m^2}\left(\frac{\dot{\mathcal{H}}}{\mathcal{H}}+\frac{2}{r_s}\right)^2\label{clvv}\\
		&\times\int_{0}^{\infty}{dk}{k^{n_s}}\left[\left(1-\frac{a}{(1+\eta)^2}\frac{\partial\eta}{\partial a}\right)GQ+\frac{a}{1+\eta}\frac{\partial(GQ)}{\partial a}\right]^2(j'_{\ell}(kr_s))^2T^2(k)\,,\nonumber
		\end{align}
while its cross-correlation with density perturbations is
		\begin{align}
		C_{\ell}^{VD}&=-\frac{8A_{s}{\tau}^{n_s-1}_0a^2}{9\pi H^4_0\Omega_m^2}\left(\frac{\dot{\mathcal{H}}}{\mathcal{H}}+\frac{2}{r_s}\right)\times\label{VD}\\
		&\int_{0}^{\infty}{dk}{k^{1+n_s}}\left[\left(1-\frac{a}{(1+\eta)^2}\frac{\partial\eta}{\partial a}\right)GQ+\frac{a}{1+\eta}\frac{\partial(GQ)}{\partial a}\right]j'_{\ell}(kr_s)j_{\ell}(kr_s)\frac{GT^2(k)}{(1+\eta)}\,.\nonumber
		\end{align}
\end{subequations}	
\paragraph{Gravitational lensing:}
Using the Limber approximation~\cite{Limber}, the autocorrelation of weak lensing is
\begin{subequations}
		\begin{align}
		C_{\ell}^{LL}=&\frac{A_{s}{\tau}_0^{n_s-1}\ell^2(\ell+1)^2}{r^2_s\nu^{4-n_s}}\int_{0}^{r_s}drT^2\left(\frac{\nu}{r}\right)\frac{(r_s-r)^2}{r^{n_s}}G^2\left(r,\frac{\nu}{r}\right)Q^2\left(r,\frac{\nu}{r}\right)\nonumber\\
		&\times \left(\frac{2+\eta(r,\nu/r)}{1+\eta(r,\nu/r)}\right)^2\,,
		\end{align}
and its cross-correlation with density becomes
		\begin{align}
		C_{\ell}^{LD}=&-\frac{8A_{s}({\tau}_0\nu)^{n_s-1}\ell(\ell+1)a}{3r_s\Omega_mH^2_0}\sqrt{\frac{\nu}{2\pi}}\int_{0}^{r_s}dr\frac{(r_s-r)}{r^{2+n_s}}\frac{j_{\ell}\left(\frac{\nu r_s}{r}\right)G(a,\nu/r)}{[1+\eta(a,\nu/r)]}\nonumber\\
		&\times\left[G\left(r,\frac{\nu}{r}\right)Q\left(r,\frac{\nu}{r}\right)\left(1+\frac{1}{1+\eta(\nu/r)}\right)\right]T^2\left(\frac{\nu}{r}\right)\,.\label{eq:ClLD}
		\end{align}
\end{subequations}
\subsection{Relativistic Effects}
We recall here that the relativistic effects are suppressed by $(\mathcal{H}/k)^2$ with respect to the leading standard effects, but they can become important at horizon scales.

\paragraph{Local potentials:}
The autocorrelation of the local potential terms is
\begin{subequations}
		\begin{align}
		C_{\ell}^{\text{lp}}&=\frac{2A_{s}{\tau}^{n_s-1}_0}{\pi}\int_{0}^{\infty}{dk}{k^{n_s-2}}j^2_{\ell}(kr_s)T^2(k)\label{PP}\\
		&\times\left[GQ\left(\frac{\dot{\mathcal{H}}}{\mathcal{H}^2}+\frac{2}{r_s\mathcal{H}}+\frac{2+\eta}{1+\eta}-\frac{a}{(1+\eta)^2}\frac{\partial\eta}{\partial a}\right)+\frac{a}{1+\eta}\frac{\partial(GQ)}{\partial a}\right]^2\,,\nonumber
		\end{align} 
and its cross-correlation with density is
		\begin{align}
		C_{\ell}^{\text{lp}D}&=-\frac{4A_{s}{\tau}^{n_s-1}_0}{3\pi\Omega_m H^2_0}\int_{0}^{\infty}dkk^{n_s}j^2_{\ell}(kr_s)\frac{GT^2(k)}{(1+\eta)}\\
		&\times \left[GQ\left(\frac{\dot{\mathcal{H}}}{\mathcal{H}^2}+\frac{2}{r_s\mathcal{H}}+\frac{2+\eta}{1+\eta}-\frac{a}{(1+\eta)^2}\frac{\partial\eta}{\partial a}\right)
		+\frac{a}{1+\eta}\frac{\partial(GQ)}{\partial a}\right]\,.\nonumber
		\end{align}
\end{subequations}
\paragraph{Shapiro time-delay:}

After using the Limber approximation, the autocorrelation of Shapiro time-delay is given by
\begin{subequations}
		\begin{align}
		C_{\ell}^{\text{std}}&=\frac{4A_{s}{\tau}^{n_s-1}_0}{r^2_s\nu^{4-n_s}}\int_{0}^{r_s}dr r^{2-n_s}T^2\left(\frac{\nu}{r}\right)G^2(r,\nu/r)Q^2(r,\nu/r)\left(\frac{2+\eta(r,\nu/r)}{1+\eta(r,\nu/r)}\right)^2\,.
		\end{align}
while its cross-correlation with density becomes
		\begin{align}
		C_{\ell}^{\text{std}D}=&-\frac{8A_{s}({\tau}_0\nu)^{n_s-1}a}{3r_s\Omega_mH^2_0}\sqrt{\frac{\nu}{2\pi}}\int_{0}^{r_s}\frac{dr}{r^{1+n_s}}\frac{j_{\ell}\left(\frac{\nu r_s}{r}\right) G(a,\nu/r)}{1+\eta(a,\nu/r)}\nonumber\\
		&\times\left(\frac{G(r,\nu/r)[2+\eta(r,\nu/r]}{1+\eta(r,\nu/r)}\right)T^2\left(\frac{\nu}{r}\right)\,.
		\end{align}
\end{subequations}
\paragraph{Integrated Sachs-Wolfe effect:}
Finally, the autocorrelation of the ISW effect under the Limber approximation is given by
\begin{subequations}
		\begin{align}
		C_{\ell}^{\text{isw}}&=\frac{A_{s}{\tau}_0^{n_s-1}}{\nu^{4-n_s}}\left(\frac{\dot{\mathcal{H}}}{\mathcal{H}^2}+\frac{2}{r_s\mathcal{H}}\right)^2\times\label{isw}\\
		&\int_{0}^{r_s}dr r^{3-n_s}T^2\left(\frac{\nu}{r}\right)\left\{\mathcal{H}(r)a(r)\frac{\partial}{\partial a}\left[G\left(r,\frac{\nu}{r}\right)Q\left(r,\frac{\nu}{r}\right)\left(\frac{2+\eta(r,\nu/r)}{1+\eta(r,\nu/r)}\right)\right]\right\}^2\,,\nonumber
		\end{align}
and its cross-correlation with density perturbations becomes
		\begin{align}
		C_{\ell}^{\text{isw}D}=&-\frac{4A_{s}({\tau}_0\nu)^{n_s-1}a}{3\Omega_mH^2_0}\sqrt{\frac{\nu}{2\pi}}\left(\frac{\dot{\mathcal{H}}}{\mathcal{H}^2}+\frac{2}{r_s\mathcal{H}}\right)\int_{0}^{r_s}\frac{dr}{r^{1+n_s}}T^2\left(\frac{\nu}{r}\right)\times\label{eq:CliswD}\\
		&\frac{j_{\ell}(\nu r_s/r)G(a_s,\nu/r)}{1+\eta(a_s,\nu/r)}\mathcal{H}(r)a(r)\frac{\partial}{\partial a}\left[G\left(r,\frac{\nu}{r}\right)Q\left(r,\frac{\nu}{r}\right)\left(\frac{2+\eta(r,\nu/r)}{1+\eta(r,\nu/r)}\right)\right]\,.\nonumber
		\end{align}
\end{subequations}
In the limit where $Q=1$ and $\eta=0$ the expressions for the standard effects \cref{eq:ClDD}-\cref{eq:ClLD} reduce to those reported in Appendix B of ref.~\cite{Bonvin-Durrer} for a $\Lambda$CDM cosmology considering a scale-invariant primordial power spectrum ($n_s=1$).

\bibliographystyle{JHEP}

\begin{thebibliography}{10}

\bibitem{Tegmark:2003}
{\scshape SDSS} collaboration, M.~Tegmark et~al., \emph{{The 3-D power spectrum
  of galaxies from the SDSS}},
  \href{https://doi.org/10.1086/382125}{\emph{Astrophys. J.} {\bfseries 606}
  (2004) 702} [\href{https://arxiv.org/abs/astro-ph/0310725}{{\ttfamily
  astro-ph/0310725}}].

\bibitem{Euclid}
R.~{Laureijs}, J.~{Amiaux}, S.~{Arduini}, J.~. {Augu{\`e}res}, J.~{Brinchmann},
  R.~{Cole} et~al., \emph{{Euclid Definition Study Report}}, {\emph{ArXiv
  e-prints} (2011) } [\href{https://arxiv.org/abs/1110.3193}{{\ttfamily
  1110.3193}}].

\bibitem{LSST}
D.~H. Weinberg, M.~J. Mortonson, D.~J. Eisenstein, C.~Hirata, A.~G. Riess and
  E.~Rozo, \emph{{Observational Probes of Cosmic Acceleration}},
  \href{https://doi.org/10.1016/j.physrep.2013.05.001}{\emph{Phys. Rept.}
  {\bfseries 530} (2013) 87} [\href{https://arxiv.org/abs/1201.2434}{{\ttfamily
  1201.2434}}].

\bibitem{Aghamousa:2016zmz}
{\scshape DESI} collaboration, A.~Aghamousa et~al., \emph{{The DESI Experiment
  Part I: Science,Targeting, and Survey Design}},
  \href{https://arxiv.org/abs/1611.00036}{{\ttfamily 1611.00036}}.

\bibitem{Kaiser}
N.~Kaiser, \emph{{Clustering in real space and in redshift space}}, {\emph{Mon.
  Not. Roy. Astron. Soc.} {\bfseries 227} (1987) 1}.

\bibitem{Hamilton}
A.~J.~S. Hamilton, \emph{{Measuring Omega and the real correlation function
  from the redshift correlation function}},
  \href{https://doi.org/10.1086/186264}{\emph{Astrophys. J.} {\bfseries 385}
  (1992) L5}.

\bibitem{Broadhurst}
T.~J. Broadhurst, A.~N. Taylor and J.~A. Peacock, \emph{{Mapping cluster mass
  distributions via gravitational lensing of background galaxies}},
  \href{https://doi.org/10.1086/175053}{\emph{Astrophys. J.} {\bfseries 438}
  (1995) 49} [\href{https://arxiv.org/abs/astro-ph/9406052}{{\ttfamily
  astro-ph/9406052}}].

\bibitem{Moessner}
R.~Moessner, B.~Jain and J.~V. Villumsen, \emph{{The effect of weak lensing on
  the angular correlation function of faint galaxies}},
  \href{https://doi.org/10.1046/j.1365-8711.1998.01225.x}{\emph{Mon. Not. Roy.
  Astron. Soc.} {\bfseries 294} (1998) 291}
  [\href{https://arxiv.org/abs/astro-ph/9708271}{{\ttfamily
  astro-ph/9708271}}].

\bibitem{Bonvin-WL}
C.~Bonvin, \emph{{Effect of Peculiar Motion in Weak Lensing}},
  \href{https://doi.org/10.1103/PhysRevD.78.123530}{\emph{Phys. Rev.}
  {\bfseries D78} (2008) 123530}
  [\href{https://arxiv.org/abs/0810.0180}{{\ttfamily 0810.0180}}].

\bibitem{Yoo}
J.~Yoo, A.~L. Fitzpatrick and M.~Zaldarriaga, \emph{{A New Perspective on
  Galaxy Clustering as a Cosmological Probe: General Relativistic Effects}},
  \href{https://doi.org/10.1103/PhysRevD.80.083514}{\emph{Phys. Rev.}
  {\bfseries D80} (2009) 083514}
  [\href{https://arxiv.org/abs/0907.0707}{{\ttfamily 0907.0707}}].

\bibitem{Yoo2}
J.~Yoo, \emph{{General Relativistic Description of the Observed Galaxy Power
  Spectrum: Do We Understand What We Measure?}},
  \href{https://doi.org/10.1103/PhysRevD.82.083508}{\emph{Phys. Rev.}
  {\bfseries D82} (2010) 083508}
  [\href{https://arxiv.org/abs/1009.3021}{{\ttfamily 1009.3021}}].

\bibitem{Bonvin-Durrer}
C.~Bonvin and R.~Durrer, \emph{{What galaxy surveys really measure}},
  \href{https://doi.org/10.1103/PhysRevD.84.063505}{\emph{Phys. Rev.}
  {\bfseries D84} (2011) 063505}
  [\href{https://arxiv.org/abs/1105.5280}{{\ttfamily 1105.5280}}].

\bibitem{BonvinGW}
C.~Bonvin, C.~Caprini, R.~Sturani and N.~Tamanini, \emph{{Effect of matter
  structure on the gravitational waveform}},
  \href{https://doi.org/10.1103/PhysRevD.95.044029}{\emph{Phys. Rev.}
  {\bfseries D95} (2017) 044029}
  [\href{https://arxiv.org/abs/1609.08093}{{\ttfamily 1609.08093}}].

\bibitem{Bonvin}
C.~Bonvin, \emph{{Isolating relativistic effects in large-scale structure}},
  \href{https://doi.org/10.1088/0264-9381/31/23/234002}{\emph{Class. Quant.
  Grav.} {\bfseries 31} (2014) 234002}
  [\href{https://arxiv.org/abs/1409.2224}{{\ttfamily 1409.2224}}].

\bibitem{Bonvin-Hui-Gaztanaga}
C.~Bonvin, L.~Hui and E.~Gaztanaga, \emph{{Optimising the measurement of
  relativistic distortions in large-scale structure}},
  \href{https://doi.org/10.1088/1475-7516/2016/08/021}{\emph{JCAP} {\bfseries
  1608} (2016) 021} [\href{https://arxiv.org/abs/1512.03566}{{\ttfamily
  1512.03566}}].

\bibitem{Amendola-Sapone}
L.~Amendola, M.~Kunz and D.~Sapone, \emph{{Measuring the dark side (with weak
  lensing)}}, \href{https://doi.org/10.1088/1475-7516/2008/04/013}{\emph{JCAP}
  {\bfseries 0804} (2008) 013}
  [\href{https://arxiv.org/abs/0704.2421}{{\ttfamily 0704.2421}}].

\bibitem{Sapone-Kunz}
D.~Sapone, M.~Kunz and M.~Kunz, \emph{{Fingerprinting Dark Energy}},
  \href{https://doi.org/10.1103/PhysRevD.80.083519}{\emph{Phys. Rev.}
  {\bfseries D80} (2009) 083519}
  [\href{https://arxiv.org/abs/0909.0007}{{\ttfamily 0909.0007}}].

\bibitem{Sapone}
S.~Nesseris and D.~Sapone, \emph{{Accuracy of the growth index in the presence
  of dark energy perturbations}},
  \href{https://doi.org/10.1103/PhysRevD.92.023013}{\emph{Phys. Rev.}
  {\bfseries D92} (2015) 023013}
  [\href{https://arxiv.org/abs/1505.06601}{{\ttfamily 1505.06601}}].

\bibitem{kessence}
C.~Armendariz-Picon, V.~F. Mukhanov and P.~J. Steinhardt, \emph{{A Dynamical
  solution to the problem of a small cosmological constant and late time cosmic
  acceleration}},
  \href{https://doi.org/10.1103/PhysRevLett.85.4438}{\emph{Phys. Rev. Lett.}
  {\bfseries 85} (2000) 4438}
  [\href{https://arxiv.org/abs/astro-ph/0004134}{{\ttfamily
  astro-ph/0004134}}].

\bibitem{ArmendarizPicon:2000ah}
C.~Armendariz-Picon, V.~F. Mukhanov and P.~J. Steinhardt, \emph{{Essentials of
  k essence}}, \href{https://doi.org/10.1103/PhysRevD.63.103510}{\emph{Phys.
  Rev.} {\bfseries D63} (2001) 103510}
  [\href{https://arxiv.org/abs/astro-ph/0006373}{{\ttfamily
  astro-ph/0006373}}].

\bibitem{Chiba:1999ka}
T.~Chiba, T.~Okabe and M.~Yamaguchi, \emph{{Kinetically driven quintessence}},
  \href{https://doi.org/10.1103/PhysRevD.62.023511}{\emph{Phys. Rev.}
  {\bfseries D62} (2000) 023511}
  [\href{https://arxiv.org/abs/astro-ph/9912463}{{\ttfamily
  astro-ph/9912463}}].

\bibitem{Bonvin:2006vc}
C.~Bonvin, C.~Caprini and R.~Durrer, \emph{{A no-go theorem for k-essence dark
  energy}}, \href{https://doi.org/10.1103/PhysRevLett.97.081303}{\emph{Phys.
  Rev. Lett.} {\bfseries 97} (2006) 081303}
  [\href{https://arxiv.org/abs/astro-ph/0606584}{{\ttfamily
  astro-ph/0606584}}].

\bibitem{Babichev:2007dw}
E.~Babichev, V.~Mukhanov and A.~Vikman, \emph{{k-Essence, superluminal
  propagation, causality and emergent geometry}},
  \href{https://doi.org/10.1088/1126-6708/2008/02/101}{\emph{JHEP} {\bfseries
  02} (2008) 101} [\href{https://arxiv.org/abs/0708.0561}{{\ttfamily
  0708.0561}}].

\bibitem{Ma:1995ey}
C.-P. Ma and E.~Bertschinger, \emph{{Cosmological perturbation theory in the
  synchronous and conformal Newtonian gauges}},
  \href{https://doi.org/10.1086/176550}{\emph{Astrophys. J.} {\bfseries 455}
  (1995) 7} [\href{https://arxiv.org/abs/astro-ph/9506072}{{\ttfamily
  astro-ph/9506072}}].

\bibitem{Bardeen}
J.~M. Bardeen, \emph{{Gauge Invariant Cosmological Perturbations}},
  \href{https://doi.org/10.1103/PhysRevD.22.1882}{\emph{Phys. Rev.} {\bfseries
  D22} (1980) 1882}.

\bibitem{Duniya:2015nva}
D.~G.~A. Duniya, D.~Bertacca and R.~Maartens, \emph{{Probing the imprint of
  interacting dark energy on very large scales}},
  \href{https://doi.org/10.1103/PhysRevD.91.063530}{\emph{Phys. Rev.}
  {\bfseries D91} (2015) 063530}
  [\href{https://arxiv.org/abs/1502.06424}{{\ttfamily 1502.06424}}].

\bibitem{Audren:2014bca}
B.~Audren, J.~Lesgourgues, G.~Mangano, P.~D. Serpico and T.~Tram,
  \emph{{Strongest model-independent bound on the lifetime of Dark Matter}},
  \href{https://doi.org/10.1088/1475-7516/2014/12/028}{\emph{JCAP} {\bfseries
  1412} (2014) 028} [\href{https://arxiv.org/abs/1407.2418}{{\ttfamily
  1407.2418}}].

\bibitem{Kunz:2006wc}
M.~Kunz and D.~Sapone, \emph{{Crossing the Phantom Divide}},
  \href{https://doi.org/10.1103/PhysRevD.74.123503}{\emph{Phys. Rev.}
  {\bfseries D74} (2006) 123503}
  [\href{https://arxiv.org/abs/astro-ph/0609040}{{\ttfamily
  astro-ph/0609040}}].

\bibitem{Koyama:2015vza}
K.~Koyama, \emph{{Cosmological Tests of Modified Gravity}},
  \href{https://doi.org/10.1088/0034-4885/79/4/046902}{\emph{Rept. Prog. Phys.}
  {\bfseries 79} (2016) 046902}
  [\href{https://arxiv.org/abs/1504.04623}{{\ttfamily 1504.04623}}].

\bibitem{DGP}
G.~R. Dvali, G.~Gabadadze and M.~Porrati, \emph{{4-D gravity on a brane in 5-D
  Minkowski space}},
  \href{https://doi.org/10.1016/S0370-2693(00)00669-9}{\emph{Phys. Lett.}
  {\bfseries B485} (2000) 208}
  [\href{https://arxiv.org/abs/hep-th/0005016}{{\ttfamily hep-th/0005016}}].

\bibitem{Koyama:2005kd}
K.~Koyama and R.~Maartens, \emph{{Structure formation in the dgp cosmological
  model}}, \href{https://doi.org/10.1088/1475-7516/2006/01/016}{\emph{JCAP}
  {\bfseries 0601} (2006) 016}
  [\href{https://arxiv.org/abs/astro-ph/0511634}{{\ttfamily
  astro-ph/0511634}}].

\bibitem{Hu-Sawicki}
W.~Hu and I.~Sawicki, \emph{{Models of f(R) Cosmic Acceleration that Evade
  Solar-System Tests}},
  \href{https://doi.org/10.1103/PhysRevD.76.064004}{\emph{Phys. Rev.}
  {\bfseries D76} (2007) 064004}
  [\href{https://arxiv.org/abs/0705.1158}{{\ttfamily 0705.1158}}].

\bibitem{Saltas:2010tt}
I.~D. Saltas and M.~Kunz, \emph{{Anisotropic stress and stability in modified
  gravity models}},
  \href{https://doi.org/10.1103/PhysRevD.83.064042}{\emph{Phys. Rev.}
  {\bfseries D83} (2011) 064042}
  [\href{https://arxiv.org/abs/1012.3171}{{\ttfamily 1012.3171}}].

\bibitem{Linde:1981}
A.~D. Linde, \emph{{A New Inflationary Universe Scenario: A Possible Solution
  of the Horizon, Flatness, Homogeneity, Isotropy and Primordial Monopole
  Problems}}, \href{https://doi.org/10.1016/0370-2693(82)91219-9}{\emph{Phys.
  Lett.} {\bfseries 108B} (1982) 389}.

\bibitem{Linder}
E.~V. Linder and R.~N. Cahn, \emph{{Parameterized Beyond-Einstein Growth}},
  \href{https://doi.org/10.1016/j.astropartphys.2007.09.003}{\emph{Astropart.
  Phys.} {\bfseries 28} (2007) 481}
  [\href{https://arxiv.org/abs/astro-ph/0701317}{{\ttfamily
  astro-ph/0701317}}].

\bibitem{Hu-viscosity}
W.~Hu, \emph{{Structure formation with generalized dark matter}},
  \href{https://doi.org/10.1086/306274}{\emph{Astrophys. J.} {\bfseries 506}
  (1998) 485} [\href{https://arxiv.org/abs/astro-ph/9801234}{{\ttfamily
  astro-ph/9801234}}].

\bibitem{DS-Finger3}
D.~Sapone and E.~Majerotto, \emph{{Fingerprinting Dark Energy III: distinctive
  marks of viscosity}},
  \href{https://doi.org/10.1103/PhysRevD.85.123529}{\emph{Phys. Rev.}
  {\bfseries D85} (2012) 123529}
  [\href{https://arxiv.org/abs/1203.2157}{{\ttfamily 1203.2157}}].

\bibitem{Renk}
J.~Renk, M.~Zumalacarregui and F.~Montanari, \emph{{Gravity at the horizon: on
  relativistic effects, CMB-LSS correlations and ultra-large scales in
  Horndeski's theory}},
  \href{https://doi.org/10.1088/1475-7516/2016/07/040}{\emph{JCAP} {\bfseries
  1607} (2016) 040} [\href{https://arxiv.org/abs/1604.03487}{{\ttfamily
  1604.03487}}].

\bibitem{Duniya:2019}
D.~Duniya, T.~Moloi, C.~Clarkson, J.~Larena, R.~Maartens, B.~Mongwane et~al.,
  \emph{{Probing beyond-Horndeski gravity on ultra-large scales}},
  \href{https://arxiv.org/abs/1902.09919}{{\ttfamily 1902.09919}}.

\bibitem{Vainshtein:1972sx}
A.~I. Vainshtein, \emph{{To the problem of nonvanishing gravitation mass}},
  \href{https://doi.org/10.1016/0370-2693(72)90147-5}{\emph{Phys. Lett.}
  {\bfseries 39B} (1972) 393}.

\bibitem{Hui:2009kc}
L.~Hui, A.~Nicolis and C.~Stubbs, \emph{{Equivalence Principle Implications of
  Modified Gravity Models}},
  \href{https://doi.org/10.1103/PhysRevD.80.104002}{\emph{Phys. Rev.}
  {\bfseries D80} (2009) 104002}
  [\href{https://arxiv.org/abs/0905.2966}{{\ttfamily 0905.2966}}].

\bibitem{Ezquiaga17}
J.~M. Ezquiaga and M.~Zumalac\'arregui, \emph{{Dark Energy After GW170817: Dead
  Ends and the Road Ahead}},
  \href{https://doi.org/10.1103/PhysRevLett.119.251304}{\emph{Phys. Rev. Lett.}
  {\bfseries 119} (2017) 251304}
  [\href{https://arxiv.org/abs/1710.05901}{{\ttfamily 1710.05901}}].

\bibitem{Sakstein:2017xjx}
J.~Sakstein and B.~Jain, \emph{{Implications of the Neutron Star Merger
  GW170817 for Cosmological Scalar-Tensor Theories}},
  \href{https://doi.org/10.1103/PhysRevLett.119.251303}{\emph{Phys. Rev. Lett.}
  {\bfseries 119} (2017) 251303}
  [\href{https://arxiv.org/abs/1710.05893}{{\ttfamily 1710.05893}}].

\bibitem{Planck18}
{\scshape Planck} collaboration, N.~Aghanim et~al., \emph{{Planck 2018 results.
  VI. Cosmological parameters}},
  \href{https://arxiv.org/abs/1807.06209}{{\ttfamily 1807.06209}}.

\bibitem{CAMB}
A.~Lewis, A.~Challinor and A.~Lasenby, \emph{{Efficient computation of CMB
  anisotropies in closed FRW models}},
  \href{https://doi.org/10.1086/309179}{\emph{Astrophys. J.} {\bfseries 538}
  (2000) 473} [\href{https://arxiv.org/abs/astro-ph/9911177}{{\ttfamily
  astro-ph/9911177}}].

\bibitem{Duniya:2013eta}
D.~Duniya, D.~Bertacca and R.~Maartens, \emph{{Clustering of quintessence on
  horizon scales and its imprint on HI intensity mapping}},
  \href{https://doi.org/10.1088/1475-7516/2013/10/015}{\emph{JCAP} {\bfseries
  1310} (2013) 015} [\href{https://arxiv.org/abs/1305.4509}{{\ttfamily
  1305.4509}}].

\bibitem{Huterer-Growth}
A.~Cooray, D.~Huterer and D.~Baumann, \emph{{Growth rate of large scale
  structure as a powerful probe of dark energy}},
  \href{https://doi.org/10.1103/PhysRevD.69.027301}{\emph{Phys. Rev.}
  {\bfseries D69} (2004) 027301}
  [\href{https://arxiv.org/abs/astro-ph/0304268}{{\ttfamily
  astro-ph/0304268}}].

\bibitem{Amendola:2017orw}
L.~Amendola, M.~Kunz, I.~D. Saltas and I.~Sawicki, \emph{{Fate of Large-Scale
  Structure in Modified Gravity After GW170817 and GRB170817A}},
  \href{https://doi.org/10.1103/PhysRevLett.120.131101}{\emph{Phys. Rev. Lett.}
  {\bfseries 120} (2018) 131101}
  [\href{https://arxiv.org/abs/1711.04825}{{\ttfamily 1711.04825}}].

\bibitem{Ligo-2017}
{\scshape GROND, SALT Group, OzGrav, DFN, INTEGRAL, Virgo, Insight-Hxmt, MAXI
  Team, Fermi-LAT, J-GEM, RATIR, IceCube, CAASTRO, LWA, ePESSTO, GRAWITA,
  RIMAS, SKA South Africa/MeerKAT, H.E.S.S., 1M2H Team, IKI-GW Follow-up, Fermi
  GBM, Pi of Sky, DWF (Deeper Wider Faster Program), Dark Energy Survey,
  MASTER, AstroSat Cadmium Zinc Telluride Imager Team, Swift, Pierre Auger,
  ASKAP, VINROUGE, JAGWAR, Chandra Team at McGill University, TTU-NRAO, GROWTH,
  AGILE Team, MWA, ATCA, AST3, TOROS, Pan-STARRS, NuSTAR, ATLAS Telescopes,
  BOOTES, CaltechNRAO, LIGO Scientific, High Time Resolution Universe Survey,
  Nordic Optical Telescope, Las Cumbres Observatory Group, TZAC Consortium,
  LOFAR, IPN, DLT40, Texas Tech University, HAWC, ANTARES, KU, Dark Energy
  Camera GW-EM, CALET, Euro VLBI Team, ALMA} collaboration, B.~P. Abbott
  et~al., \emph{{Multi-messenger Observations of a Binary Neutron Star
  Merger}}, \href{https://doi.org/10.3847/2041-8213/aa91c9}{\emph{Astrophys.
  J.} {\bfseries 848} (2017) L12}
  [\href{https://arxiv.org/abs/1710.05833}{{\ttfamily 1710.05833}}].

\bibitem{Nersisyan:2018auj}
H.~Nersisyan, N.~A. Lima and L.~Amendola, \emph{{Gravitational wave speed:
  Implications for models without a mass scale}},
  \href{https://arxiv.org/abs/1801.06683}{{\ttfamily 1801.06683}}.

\bibitem{Saltas:2014dha}
I.~D. Saltas, I.~Sawicki, L.~Amendola and M.~Kunz, \emph{{Anisotropic Stress as
  a Signature of Nonstandard Propagation of Gravitational Waves}},
  \href{https://doi.org/10.1103/PhysRevLett.113.191101}{\emph{Phys. Rev. Lett.}
  {\bfseries 113} (2014) 191101}
  [\href{https://arxiv.org/abs/1406.7139}{{\ttfamily 1406.7139}}].

\bibitem{Seljak:2009}
U.~Seljak, \emph{{Extracting primordial non-gaussianity without cosmic
  variance}}, \href{https://doi.org/10.1103/PhysRevLett.102.021302}{\emph{Phys.
  Rev. Lett.} {\bfseries 102} (2009) 021302}
  [\href{https://arxiv.org/abs/0807.1770}{{\ttfamily 0807.1770}}].

\bibitem{Yoo-Seljak:2012}
J.~Yoo, N.~Hamaus, U.~c.~v. Seljak and M.~Zaldarriaga, \emph{Going beyond the
  kaiser redshift-space distortion formula: A full general relativistic account
  of the effects and their detectability in galaxy clustering},
  \href{https://doi.org/10.1103/PhysRevD.86.063514}{\emph{Phys. Rev. D}
  {\bfseries 86} (2012) 063514}.

\bibitem{Linder:2005in}
E.~V. Linder, \emph{{Cosmic growth history and expansion history}},
  \href{https://doi.org/10.1103/PhysRevD.72.043529}{\emph{Phys. Rev.}
  {\bfseries D72} (2005) 043529}
  [\href{https://arxiv.org/abs/astro-ph/0507263}{{\ttfamily
  astro-ph/0507263}}].

\bibitem{Narikawa:2009ux}
T.~Narikawa and K.~Yamamoto, \emph{{Characterising linear growth rate of
  cosmological density perturbations in f(R) model}},
  \href{https://doi.org/10.1103/PhysRevD.81.129903,
  10.1103/PhysRevD.81.043528}{\emph{Phys. Rev.} {\bfseries D81} (2010) 043528}
  [\href{https://arxiv.org/abs/0912.1445}{{\ttfamily 0912.1445}}].

\bibitem{Basilakos:2012uu}
S.~Basilakos and A.~Pouri, \emph{{The growth index of matter perturbations and
  modified gravity}},
  \href{https://doi.org/10.1111/j.1365-2966.2012.21168.x}{\emph{Mon. Not. Roy.
  Astron. Soc.} {\bfseries 423} (2012) 3761}
  [\href{https://arxiv.org/abs/1203.6724}{{\ttfamily 1203.6724}}].

\bibitem{MGCAMB}
A.~Hojjati, L.~Pogosian and G.-B. Zhao, \emph{{Testing gravity with CAMB and
  CosmoMC}}, \href{https://doi.org/10.1088/1475-7516/2011/08/005}{\emph{JCAP}
  {\bfseries 1108} (2011) 005}
  [\href{https://arxiv.org/abs/1106.4543}{{\ttfamily 1106.4543}}].

\bibitem{Limber}
M.~LoVerde and N.~Afshordi, \emph{{Extended Limber Approximation}},
  \href{https://doi.org/10.1103/PhysRevD.78.123506}{\emph{Phys. Rev.}
  {\bfseries D78} (2008) 123506}
  [\href{https://arxiv.org/abs/0809.5112}{{\ttfamily 0809.5112}}].

\end{thebibliography}

\providecommand{\href}[2]{#2}\begingroup\raggedright\endgroup

\end{document}